\documentclass[pra,twocolumn,superscriptaddress]{revtex4-2}
\usepackage{multirow}
\usepackage{rotating}
\usepackage{booktabs,tabularray}
\usepackage{amsmath}
\usepackage{mathrsfs}
\usepackage[protrusion=true,expansion=true]{microtype}
\usepackage{times}
\usepackage{bm,xcolor,braket, tikz}
\usepackage{graphicx}
\usepackage[normalem]{ulem}
\renewcommand\sout{\bgroup\markoverwith{\textcolor{magenta}{\rule[0.5ex]{2pt}{2pt}}}\ULon}

\usepackage{hyperref}
\hypersetup{colorlinks,linkcolor={blue},citecolor={blue},urlcolor={blue}}
\usepackage{amssymb,physics}
\usepackage[utf8]{inputenc}
\usepackage[english]{babel}
\usepackage{bm}
\newcommand{\Ltilde}{{\cal T}}
\makeatletter
\def\graphicscale{\twocolumn@sw{0.3}{0.4}}
\def\graphicthreescale{\twocolumn@sw{0.3}{0.4}}

\newcommand{\be}{\begin{equation}}
\newcommand{\ee}{\end{equation}}
\newcommand{\bea}{\begin{align}}
\newcommand{\eea}{\end{align}}
\newcommand{\de}{\partial}

\usepackage{float}

\usepackage{dsfont}
\usepackage{mathrsfs}
\usepackage{comment}

\begin{document}
\title{One-body correlations and momentum distributions of trapped one-dimensional\\ Bose gases at finite temperature}

\author{Attila Tak\'acs}
\altaffiliation[Attila Tak\'acs and Yicheng Zhang~contributed equally to this work.]{}
\affiliation{Universit\'e de Lorraine, CNRS, LPCT, F-54000 Nancy, France}
\affiliation{SISSA and INFN, via Bonomea 265, 34136 Trieste, Italy}
\author{Yicheng Zhang}
\altaffiliation[Attila Tak\'acs and Yicheng Zhang~contributed equally to this work.]{}
\affiliation{Homer L. Dodge Department of Physics and Astronomy, The University of Oklahoma, Norman, OK 73019, USA}
\affiliation{Center for Quantum Research and Technology, The University of Oklahoma, Norman, OK 73019, USA}
\author{Pasquale Calabrese}
\affiliation{SISSA and INFN, via Bonomea 265, 34136 Trieste, Italy}
\affiliation{International Centre for Theoretical Physics (ICTP), Strada Costiera 11, 34151 Trieste, Italy}
\author{Jer\^ome Dubail}
\affiliation{Universit\'e de Lorraine, CNRS, LPCT, F-54000 Nancy, France}
\affiliation{CESQ and ISIS (UMR 7006), University of Strasbourg and CNRS, 67000 Strasbourg, France}
\author{Marcos Rigol}
\affiliation{Department of Physics, The Pennsylvania State University, University Park, PA 16802, USA}
\author{Stefano Scopa}
\affiliation{SISSA and INFN, via Bonomea 265, 34136 Trieste, Italy}
\affiliation{Laboratoire de Physique Th\'eorique et Mod\'elisation, CNRS UMR 8089, CY Cergy Paris Universit\'e, 95302 Cergy-Pontoise Cedex, France}
\affiliation{Laboratoire de Physique de l’\'Ecole Normale Superieure, CNRS, ENS \& Universit\'e PSL,
Sorbonne Universit\'e, Universit\'e Paris Cit\'e, 75005 Paris, France}

\date{\today}

\begin{abstract}
We introduce a general approximate method for calculating the one-body correlations and the momentum distributions of one-dimensional Bose gases at finite interaction strengths and temperatures trapped in smooth confining potentials. Our method combines asymptotic techniques for the long-distance behavior of the gas (similar to Luttinger liquid theory) with known short-distance expansions. We derive analytical results for the limiting cases of strong and weak interactions and provide a general procedure for calculating one-body correlations at any interaction strength. A step-by-step explanation of the numerical method used to compute Green’s functions (needed as input to our theory) is included. We benchmark our method against exact numerical calculations and compare its predictions to recent experimental results.
\end{abstract}

\maketitle

\section{Introduction}\label{sec:intro}

Since the early days of Bose-Einstein condensation in ultracold gas experiments~\cite{davis1995bose, anderson1995observation}, the momentum distribution of the atoms has been a pivotal experimental observable~\cite{bloch2008review}. Measured via time-of-flight imaging, the momentum distribution has allowed to observe and characterize a wide range of phenomena in a wide range of systems~\cite{davis1995bose, anderson1995observation, bloch2008review, bourdel2003measurement, regal2005momentum, kinoshita2006quantum, stewart2010verification}. In recent years, momentum distribution measurements in ultracold gases in one-dimensional (1D) and close-to-1D geometries have allowed to observe dynamical fermionization during the expansion in 1D~\cite{wilson2020observation}, test the accuracy of generalized hydrodynamics~\cite{malvania2021generalized}, study the 2D-1D crossover~\cite{guo20231d2dcross}, probe the effect of dipolar interactions in 1D gases~\cite{li2023dipolar}, observe hydrodynamization after Bragg scattering pulses~\cite{le2023observation}, unveil cooling by dimensional reduction~\cite{li2023dipolar, guo2024cooling}, and characterize the dynamics of dipolar-interaction stabilized many-body quantum scars~\cite{yang2023phantom}. 

The momentum distribution $f({\bf p})$ is the Fourier transform,
\begin{equation}
f({\bf p})=\int d {\bf x}  \int d{\bf y} \; e^{i {\bf p} ({\bf x}-{\bf y})} g_1({\bf x}, {\bf y}) ,
\label{eq:MDdef}
\end{equation}
of the equal-time correlation function
\begin{equation}
    \label{eq:g1_def}
    g_1({\bf x}, {\bf y}) = \langle \hat{\Psi}^\dagger({\bf x}) \hat{\Psi}({\bf y}) \rangle,
\end{equation}
which is known as the one-body density matrix (OBDM). Here $\hat{\Psi}^\dagger({\bf x})$ and $\hat{\Psi}({\bf x})$ are the one-particle creation and annihilation operators, respectively, at position ${\bf x}$. We set $\hbar=1$ throughout our analytical derivations, and reintroduce $\hbar$ when comparing the analytical and numerical results in Sec.~\ref{sec:LLvsQMC}. 

Nonlocal correlation functions like $g_1({\bf x},{\bf y})$ are generally challenging to compute both analytically and numerically. Consequently, predicting the momentum distribution theoretically is difficult, specially in correlated gases out of equilibrium~\cite{kinoshita2006quantum, wilson2020observation, malvania2021generalized, le2023observation, yang2023phantom}. In the context of bosonic gases, this challenge has attracted significant attention from theorists over the years~\cite{cazalilla2011one}. Analytical results have predominantly been restricted to 1D. Even in 1D, direct calculations for microscopic Hamiltonians are typically limited to hard-core bosons~\cite{lenard1964momentum, lenard1972some, vaidya1979one, pezer2007momentum}, which can be mapped onto noninteracting fermions through the Bose-Fermi mapping~\cite{girardeau1960relationship}. Numerically, the momentum distribution in equilibrium can be obtained using quantum Monte Carlo simulations~\cite{jacqmin2012momentum, fang2016momentum, xu2015universal} or, for integrable gases, by using sophisticated form-factor resummation methods \cite{caux2007one, caux2009correlation}. Furthermore, numerical results in and out of equilibrium can be obtained for lattice hard-core bosons at zero~\cite{rigol_muramatsu_05b, rigol_muramatsu_05c} and finite~\cite{rigol_05, xu_rigol_17} temperatures. Such lattice calculations have been used, in the low-density limit, to understand some of the recent experimental results in the continuum mentioned earlier~\cite{wilson2020observation, malvania2021generalized, li2023dipolar, le2023observation, yang2023phantom}. 

%%%%%%%%%%%%%%%%%%%%%%%%%%%
\begin{figure*}[t]
  \includegraphics[width=\textwidth]{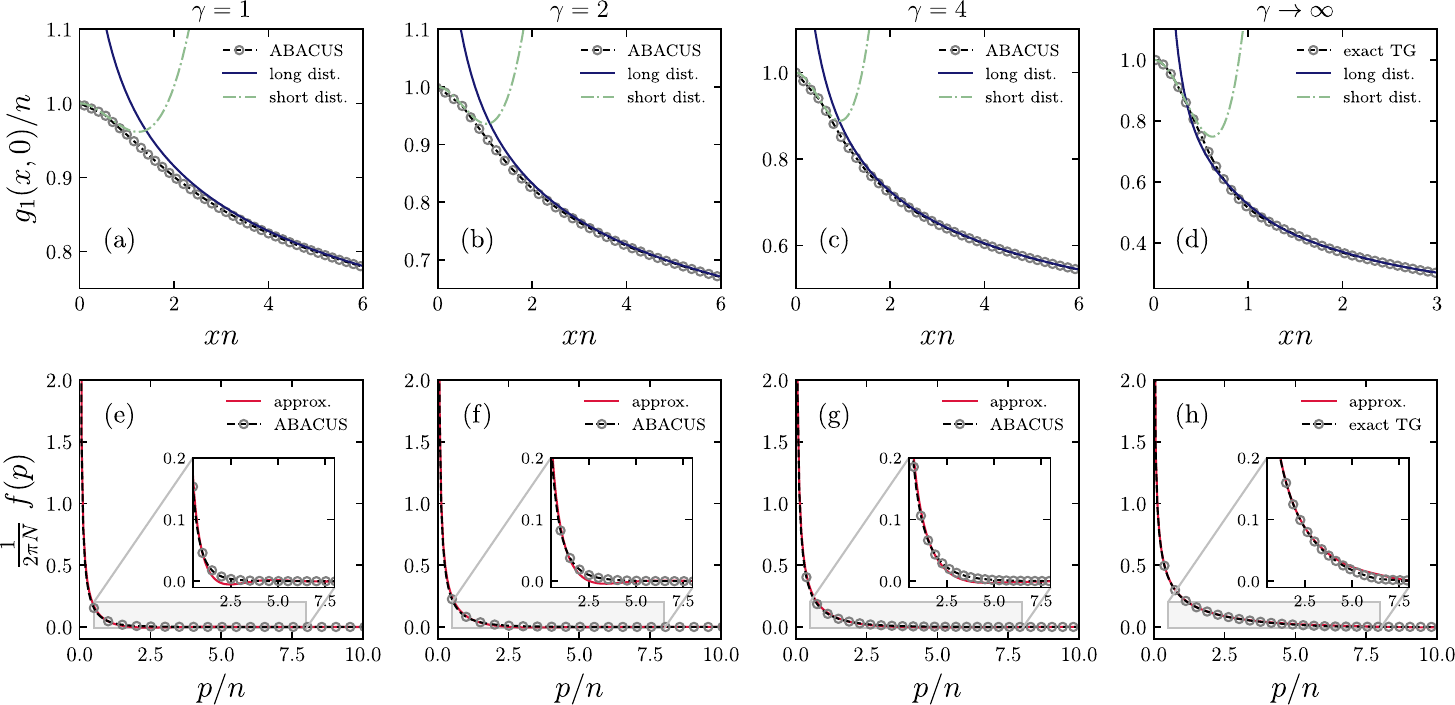}
  \vspace{-0.7cm}
  \caption{{\it Top}: Exact results for $g_1(x,0)$ (symbols, ABACUS data of Ref.~\cite{caux2007one} at finite interaction, and exact Bose-Fermi mapping result at infinite repulsion) compared against the leading order ($m=0$) of Eq.~\eqref{eq:LL_exp} [solid line, see Eq.~\eqref{eq:ll-asy}] and to Eq.~\eqref{eq:short_dist_exp} up to $q=4$ [dot-dashed line, see Eq.~\eqref{eq:short_dist_exp-lowest}]. {\it Bottom}: Corresponding normalized momentum distribution $f(p)$ obtained from the exact results (dashed line) and from the minimum of $g_1(x,0)$ in the two asymptotic regimes shown for $g_1(x,0)$ (solid line). The insets provide a magnification of $f(p)$ at intermediate momenta. The reduced interaction coupling $\gamma$ increases from the left to right.}
  \label{fig:g1_momdist_reg}
\end{figure*}
%%%%%%%%%%%%%%%%%%%%%%%%%%%

There also exist well-known asymptotic results for the OBDM of 1D bosonic gases. In particular, in homogeneous (i.e., translational invariant) ground states, the {\it long-distance} asymptotic behavior of $g_1(x,y)$ is predicted by Luttinger liquid theory~\cite{efetov1976correlation, haldane1981luttinger, giamarchi2003quantum, cazalilla2004bosonizing} to be of the general form:
\begin{equation}
   \label{eq:LL_exp}
   \frac{ g_1(x,y)}{n} \underset{|x-y|\gg d(n)}{\simeq} \sum_{m \geq 0} \frac{B_m 2\cos(2mk_F|x-y|)}{(n |x-y|)^{2m^2K + \frac{1}{2K}}},
\end{equation}
where $n$ is the 1D atom density, $k_F = \pi n$ is the associated Fermi wavevector, and $K$ is the dimensionless Luttinger parameter, which depends on the strength of the interaction between the bosons. $d(n)$ is the typical length scale of the microscopic problem, thereafter taken as the maximum between the interparticle distance $n^{-1}$, and the healing length $\xi_\text{heal}\propto 1/\sqrt{n}$, namely $d(n)=\max(1/n,\xi_\text{heal})$. The sum in Eq.~\eqref{eq:LL_exp} runs over integer numbers $m\geq 0$, physically encoding low-energy momentum $\pm 2mk_F$ processes~\cite{giamarchi2003quantum}. Each of these terms comes with a dimensionless amplitude $B_m$ that depends on the microscopic details of the model~\cite{shashi2011nonuniversal,shashi2012exact}. It follows from Eq.~(\ref{eq:LL_exp}) that the momentum distribution of the ground state of an infinitely long 1D Bose gas exhibits a peak $f(p) \sim |p|^{\frac{1}{2K}-1}$ for $p\to 0$, as well as weaker singularities about integer multiples of $2k_F$, $f(p) \sim |p \mp 2m k_F|^{\frac{1}{2K} + 2m^2 K -1}$ when $p \rightarrow \pm 2 m k_F$.

Asymptotic results are also available for the {\it short-distance} behavior of $g_1(x,y)$. In particular, for the ground state of the 1D Bose gas with $\delta$-interaction, also known as the Lieb-Liniger model~\cite{lieb1963exact} (see Sec.~\ref{sec:LLreview} for details)~\cite{gangardt2004universal, olshanii2003short, Olshanii_2017}:
\begin{equation}
\label{eq:short_dist_exp}
\frac{g_1(x,y)}{n} \underset{|x - y| \ll d(n)}{\simeq} 1+ \sum_{q>1} C_q (n|x-y|)^q
\end{equation}
where the dimensionless coefficients $C_q$ can be expressed in terms of local thermodynamic quantities in the gas.

In Figs.~\ref{fig:g1_momdist_reg}(a)--\ref{fig:g1_momdist_reg}(d), we compare asymptotic results for the ground state of the repulsive Lieb-Liniger model to numerical form-factor resummation results at finite interaction and exact Bose-Fermi mapping result at infinite repulsion for $N=100$ bosons in a ring with unit density~\cite{caux2007one}. Due to the finite length $L$ of the ring, to plot the long-distance results from Eq.~\eqref{eq:LL_exp} we replace $|x-y|$ by the chord distance $L \sin(\pi |x-y|/L)/\pi$. The agreement between the numerical and the asymptotic results is excellent at short and long distances for all values of the dimensionless coupling parameter $\gamma$ (see Sec.~\ref{sec:LLreview}) shown. In the strongly interacting regime, the asymptotic results deviate from the numerical ones within a narrow window about $|x-y|\sim 1/n$. In the limit of infinite repulsion (hard-core bosons), this observation dates back to the seminal work of Vaidya and Tracy~\cite{vaidya1979one}. The deviations of the short-distance expansion for arbitrary interaction strengths were recently discussed in Ref.~\cite{DeRosi2024}. In Figs.~\ref{fig:g1_momdist_reg}(e)--\ref{fig:g1_momdist_reg}(h) we compare $f(p)$ obtained using the Fourier transform of the asymptotic results for $g_1(x,y)$ (closest to the numerical ones) to the numerical results for $f(p)$~\cite{caux2007one}. The agreement is excellent at low momentum ($p\lesssim k_F$). The disagreement becomes visible, see the insets, for $p\gtrsim k_F$ before $f(p)$ vanishes at very high momentum.

The previous observations for the homogeneous ground state motivate the analysis in this paper. We use asymptotic results for the long- and short-distance behavior of the OBDM at finite temperature (see also Refs.~\cite{DeRosi2024, DeRosi2023}) and in the presence of confining potentials to compute momentum distributions that are of direct relevance to current experiments with ultracold 1D gases (as we show in Sec.~\ref{sec:nummethod}). Our results are compared to those obtained for exactly solvable cases, such as trapped hard-core bosons at finite temperature and weakly interacting bosons in the quasicondensate regime. For intermediate interactions strengths, which are not exactly solvable, we benchmark our results against those of quantum Monte Carlo simulations.

The paper is organized as follows. In Sec.~\ref{sec:LLreview}, we introduce the Lieb-Liniger model and review known results for the long- and short-distance asymptotics of the OBDM. In Sec.~\ref{section:LDA}, we study trapped 1D gases in equilibrium at finite but low temperature using the ``inhomogeneous Luttinger liquid'' approach~\cite{cazalilla2004bosonizing, dubail2017conformal, brun2017one, brun2018inhomogeneous, Bastianello_2020, scopa2020one}. We derive analytical expressions for the OBDM in traps at finite temperature in the hard-core (Tonks-Girardeau) limit and in the weakly-interacting (Gross-Pitaevskii) limit. In Sec.~\ref{sec:nummethod}, we provide a detailed discussion of our numerical method for evaluating the OBDM in the inhomogeneous Luttinger liquid for arbitrary repulsion strengths, generalizing the method of Ref.~\cite{Bastianello_2020} to finite temperature. In Sec.~\ref{sec:nummethod}, we benchmark our approach against exact numerical calculations and compare its predictions to recent experimental results. We summarize our results and discuss potential extensions in Sec.~\ref{sec:conclusion}.

\section{Lieb-Liniger model}
\label{sec:LLreview}

Throughout this paper, we focus on 1D gases of bosons with repulsive contact interactions. In the absence of an external potential, the corresponding model was introduced and solved by Lieb and Liniger~\cite{lieb1963exact}:
\be\label{eq:LL-model}
\hat{H}_0=\frac{1}{2}\int_{-\frac L2}^{\frac L2} dx\ \hat\Psi^\dagger(x)\left[-\de^2_x+ c\hat\Psi^\dagger(x)\hat\Psi(x)\right] \hat\Psi(x) ,
\ee
where $\hat\Psi^\dagger(x)$ and $\hat\Psi(x)$ are bosonic creation and annihilation operators, respectively, at position $x$ in a ring of length $L$. We set the mass of the bosons $m=1$, and $c>0$ is the strength of the repulsive contact interaction. 

The Lieb-Liniger model can be solved using the Bethe ansatz~\cite{lieb1963exact, gaudin2014bethe, korepin1997quantum}. Focusing on the sector with $N$ bosons, the Hamiltonian~\eqref{eq:LL-model} can be written in the first-quantized form
\be
\hat {\cal H}=-\frac{1}{2}\sum_{j=1}^N \frac{\de^2}{\de x_j^2} +c \sum_{1\leq j<\ell \leq N} \delta(x_j-x_\ell) ,
\ee
with associated many-body eigenstates
\be
\hat {\cal H}\, \chi(\bm\lambda, \vec{x})=E(\bm\lambda)\,\chi(\bm\lambda,\vec{x})
\ee
$\vec{x}=(x_1,\dots,x_N)$, whose explicit expression can be found, e.g., in Refs.~\cite{lieb1963exact, gaudin2014bethe, korepin1997quantum}. Importantly, these eigenstates are labeled by a set of real spectral parameters $\bm\lambda=(\lambda_1,\dots,\lambda_N)$ (or rapidities) whose allowed values are the solutions of the Bethe equations:
\be
e^{i\lambda_j L}=\prod_{\ell\neq j} \frac{\lambda_j-\lambda_\ell+ i c}{\lambda_j-\lambda_\ell + i c}, \quad j=1,\dots,N,
\ee
or, equivalently, in logarithmic form
\begin{equation}
    \lambda_j +\frac{1}{L}\sum_{\ell=1}^N 2\arctan\frac{\lambda_j-\lambda_\ell}{c}=\frac{2\pi}{L}I_j.
    \label{eq:Betheeq}
\end{equation}
Equation~\eqref{eq:Betheeq} uniquely specifies a rapidity set $\bm\lambda$ [hence, an eigenstate of the Hamiltonian \eqref{eq:LL-model}] for a set of distinct integers (half-integers) $I_1,\,\dots,\,I_N$ for $N$ even (odd). In fact, one may interpret the r.h.s.~of Eq.~\eqref{eq:Betheeq} as the set of momenta of a noninteracting Fermi gas, and thus one can think of the corresponding rapidities as imposing a nontrivial quantization condition due to the contact interactions. For instance, the ground state set $\bm\lambda_\text{GS}$ is obtained from the equally spaced configuration $I_j=-\frac{N+1}{2}+j$, i.e., by filling a Fermi sea for the associated noninteracting system. 

In the thermodynamic limit, $N\to\infty$ and $L\to \infty$ at fixed density $n=N/L$, it is convenient to replace the rapidity set $\bm\lambda$ with a smooth density distribution $\rho(\lambda_j)=\lim_{N,L\to\infty} 1/[L(\lambda_{j+1}-\lambda_j)]$. In the ground state, the latter satisfies the following integral equation
\be
2\pi\rho(\lambda)= 1 + \int_{-\lambda_F}^{\lambda_F} d\lambda' \frac{2c  \rho(\lambda')}{c^2 + (\lambda - \lambda')^2},
\label{eq:rootdensity}
\ee
with $\lambda_F$ fixed by $n=\int_{-\lambda_F}^{\lambda_F} d\lambda \ \rho(\lambda)$. For later convenience, we also introduce the dressing operation of a generic function of rapidities $h$, defined as the solution to the integral equation~\cite{doyon_lecture_notes_GHD}
\begin{equation}
    h^{\rm dr}(\lambda) = h(\lambda) + \int_{-\lambda_F}^{\lambda_F} \frac{d\lambda'}{2\pi} \frac{2c}{c^2 + (\lambda - \lambda')^2} h^{\rm dr}( \lambda'),
    \label{eq:dressing}
\end{equation}
in terms of which $\rho(\lambda)\equiv \left[1/(2\pi)\right]^{\rm dr}$. It is also possible to express Eq.~\eqref{eq:rootdensity} in terms of dimensionless variables $\alpha=c/\lambda_F$ and $g(u)\equiv \rho(\lambda_F u)$:
\be
2\pi g(u,\alpha)=1 +\int_{-1}^1 du' \frac{2\alpha g(u',\alpha)}{\alpha^2+(u-u')^2},
\ee
normalized such that $\gamma\int_{-1}^{1} du \ g(u,\alpha)=\alpha$. As a result, one finds that the equilibrium properties of the Lieb-Liniger gas depend uniquely on the dimensionless reduced coupling 
\be
\gamma=c/n,
\ee
with $\gamma\to 0$ and $\gamma\to\infty$ corresponding to the limits of weak and strong interactions, respectively.

Although the Bethe ansatz approach provides an exact understanding of the spectrum of the Lieb-Liniger model~\eqref{eq:LL-model}, the calculation of correlation functions within this framework is a formidable challenge.  Since determinant formulas for the field form-factors $\langle\bm\lambda|\hat\Psi^\dagger(0)|\bm\mu\rangle$ between two generic Bethe-ansatz eigenstates have been determined \cite{slavnov1989}, one may express the OBDM \eqref{eq:g1_def} as
\be\label{eq:ff-exp}
g_1(x,0)=\sum_{\{\bm\mu\}} \frac{\langle\bm\lambda|\hat\Psi^\dagger(x)|\bm\mu\rangle\langle \bm\mu|\hat\Psi(0)|\bm\lambda\rangle }{\langle\bm\lambda|\bm\lambda\rangle\langle\bm\mu|\bm\mu\rangle},
\ee
where $\bm\lambda$ is the rapidity set of the reference state (e.g., the ground state). The evaluation of Eq.~\eqref{eq:ff-exp} requires the summation over the intermediate Bethe-ansatz states $\ket{\bm\mu}$, which beyond few-particle systems is a challenging task that needs to be tackled using sophisticated numerical algorithms, see, e.g., Ref.~\cite{caux2009correlation}. The numerical evaluation of Eq.~\eqref{eq:ff-exp} for the ground state, reported in Ref.~\cite{caux2007one}, yielded the results plotted in Fig.~\ref{fig:g1_momdist_reg}.

\subsection{Long-distance asymptotics of the OBDM}

Alternatively, a universal description of the system's correlations can be obtained at low energies using Luttinger liquid theory~\cite{haldane1981luttinger, giamarchi2003quantum}. The main idea of this effective low-energy theory is to encode low-energy quantum fluctuations in terms of fluctuating bosons on top of the Fermi sea $\lambda \in [-\lambda_F,\lambda_F]$ obtained through the thermodynamic Bethe ansatz. Following this approach, it is possible to write the so-called harmonic fluid expansion of the field operator~\cite{haldane1981luttinger,giamarchi2003quantum,cazalilla2004bosonizing,cazalilla2011one}
\be\label{eq:haldane}
\hat\Psi^\dagger(x)\! =\! \sqrt{n-\frac{\de_x\hat\phi}{\pi}}\!\sum_{m\in\mathbb{Z}}\!\left[ \sqrt{B_{|m|} n^{-\Delta_m}} e^{im[k_F x+\hat\phi(x)]}\right]\! e^{-i\hat\theta(x)},
\ee
with the density $\de_x\hat\phi$ and phase $\hat\theta$ fluctuating fields satisfying $[ \de_x\hat{\phi}(x), \hat\theta(x')] = -i\pi \delta(x-x')$. In Eq.~\eqref{eq:haldane}, $B_m\equiv B_m(\gamma)$ are dimensionless nonuniversal amplitudes associated to Umklapp scattering, whose values are obtained from field form factors in the thermodynamic limit, as detailed in Appendix~\ref{app:nonuniv-coef} (see also Refs.~\cite{shashi2011nonuniversal, shashi2012exact, brun2018inhomogeneous, scopa2020one}). The properties of such fluctuating bosons are determined by the Luttinger Hamiltonian
\begin{equation} \label{eq:LL_hom}
        \hat{\mathscr{H}} =\frac{1}{2 \pi } \int_0^L dx  \left( vK\left[\partial_x \hat{\theta}(x)\right]^2 + \frac{v}{K} \left[ \partial_x \hat{\phi}(x)\right]^2 \right),
\end{equation}
where $v$ is the sound velocity and $K$ is the Luttinger parameter, respectively. For the Lieb-Liniger model \eqref{eq:LL-model}, these parameters are not independent, $v=\pi n/K$, with $K=[1^{\rm dr}(\lambda_F)]^2\geq 1$ for repulsive interactions.

By establishing the two-point correlation of the fluctuating fields via Eq.~\eqref{eq:LL_hom}, the Luttinger liquid theory enables the calculation of higher-order correlators by means of Wick's theorem. For the specific case of $g_1(x,y)$, this leads to Eq.~\eqref{eq:LL_exp}~\cite{giamarchi2003quantum, cazalilla2004bosonizing, cazalilla2011one}. When evaluating the sum in Eq.~\eqref{eq:LL_exp}, one can use the fact that each harmonic $m$ contributes to the expansion in Eq.~\eqref{eq:haldane} as a short-distance correction that scales like $n^{-\Delta_m}$, with $\Delta_m=2m^2 K + \frac{1}{2K}$~\cite{giamarchi2003quantum}. Hence, by truncating the sum at its leading order ($m=0$), one obtains the long-distance asymptotics of the OBDM [valid for $|x| \gg d(n)$]
\begin{equation}\label{eq:ll-asy}
   \frac{ g_1(x,0)}{n} \approx  \frac{B_0 n^{-\frac{1}{2K}}}{|x|^{\frac{1}{2K}}},
\end{equation} 
with $|x|$ replaced by $L \sin(\pi|x|/L)/\pi$ in finite-size systems. Furthermore, the Luttinger liquid theory allows one to account for small thermal fluctuations. By incorporating the effect of a finite temperature $T$ in the low-energy description of fluctuating fields, one obtains:
\begin{equation}\label{eq:LL-thermal-hom}
   \frac{ g_1(x,0)}{n} \approx  \frac{B_0 n^{-\frac{1}{2K}}}{\left[\xi_T\sinh\left(\frac{|x|}{\xi_T}\right) \right]^{\frac{1}{2K}}},
\end{equation}
valid for a translationally invariant gas when $\xi_T\ll  d(n) \ll |x|$, with thermal length $\xi_T=\pi T/v$ (we set the Boltzmann constant $k_B=1$). See, e.g., Secs.~3 and 4 of Ref.~\cite{giamarchi2003quantum} or the appendices of Ref.~\cite{cazalilla2004bosonizing} for a derivation of Eq.~\eqref{eq:LL-thermal-hom}.

\subsection{Short-distance asymptotics of the OBDM}

The long-distance asymptotics in Eq.~\eqref{eq:LL_exp} exhibits an ultraviolet divergence in the limit $|x-y|\to 0$. This singularity, absent in the microscopic model, is inherent to the Luttinger liquid description and must be regularized in order to obtain the momentum distribution via Fourier transform [cf.~Eq.~\eqref{eq:MDdef}]. To this end, the short-distance expansion of $g_1(x,y)$ for $|x-y|\ll d(n)$, reported in Eq.~\eqref{eq:short_dist_exp} and derived in Refs.~\cite{gangardt2004universal, olshanii2003short, Olshanii_2017}, can be used. Retaining only the lowest orders, one has
\begin{equation}
\frac{g_1(x,0)}{n}\overset{x\to y}{\approx} 1+C_2(n|x|)^2 +C_3(n|x|)^3+C_4(n|x|)^4,
\label{eq:short_dist_exp-lowest}
\end{equation}
with the coefficients expressed in terms of thermodynamic Bethe ansatz quantities
\begin{equation}
C_2(\gamma)=-\frac{\epsilon_2(\gamma)-\gamma \epsilon_2'(\gamma)}{2} ,
\end{equation} 
\begin{equation}
C_3(\gamma)=\frac{\gamma^2 \epsilon_2'(\gamma)}{12},
\end{equation}
and
\begin{equation}
    C_4(\gamma)=\frac{18 \epsilon_2^2 -9\epsilon_4 -2\epsilon_2(2+3\epsilon_2')\gamma + \gamma[2\epsilon_4' + \epsilon_2' \gamma(2+\gamma)]}{24}.
\end{equation}
Here we defined the functions 
\be
\epsilon_{2k}(\gamma)=\frac{\int_{-1}^1 du \ g(u,\alpha_\gamma) u^{2k}}{\left[\int_{-1}^1 du \ g(u,\alpha_\gamma)]\right]^{2k+1}}.
\ee
Strictly speaking, Eq.\eqref{eq:short_dist_exp-lowest} is valid for the ground state of the Lieb-Liniger model. However, while finite but low temperatures qualitatively change the long-distance decay of the OBDM from algebraic to exponential [cf.~Eqs.~\eqref{eq:ll-asy} and~\eqref{eq:LL-thermal-hom}], they do not significantly affect its short-distance behavior. The validity of Eq.~\eqref{eq:short_dist_exp-lowest} for the finite-temperature gas was tested against quantum Monte Carlo simulations in Refs.~\cite{DeRosi2023,DeRosi2024}.

As discussed in the introduction, Fig.~\ref{fig:g1_momdist_reg} shows the ground-state OBDM for a translationally-invariant gas of size $L$, obtained by combining the asymptotic results of Eqs.~\eqref{eq:ll-asy} and~\eqref{eq:short_dist_exp-lowest}. For the long-distance behavior, we retained only the leading-order term ($m=0$), although we verified that subleading corrections ($m=\pm1$) do not significantly affect the momentum distribution within the range of momenta that are shown. The matching at intermediate scales $|x-y|\sim d(n)$ is done by taking the minimum of the two asymptotic curves. While improved results could be obtained through, e.g., a polynomial interpolation between the two asymptotic regimes, we find that our approach provides accurate results without the need of further manipulations.  

\section{Trapped 1D Bose gases}
\label{section:LDA}

In experiments with ultracold bosonic gases in 1D geometries, which can be realized using 2D optical lattices~\cite{bloch2008review, cazalilla2011one} or atom chips~\cite{folman2002atomchip}, a confining potential $V(x)$ is present, so the corresponding 1D gases are modeled using the Hamiltonian:
\be\label{eq:LL-model-trap}
\hat{H}=\hat{H}_0+\int_{-\frac L2}^{\frac L2} dx\ V(x) \hat\Psi^\dagger(x)\hat\Psi(x).
\ee
The confining potential breaks the Bethe ansatz solvability of the model. Yet, one can use the local density approximation (LDA) to describe local quantities in the inhomogeneous system using the corresponding Bethe ansatz results for the homogeneous gas with a local chemical potential $\mu - V(x)$, see, e.g., Refs.~\cite{bloch2008review, cazalilla2011one}. This simple approach provides an accurate description of inhomogeneous gases whenever the length scale associated to the changes in the density due to the external potential is much longer than the interparticle distance, namely, whenever $n^{-1}(x)\ll |\de_x \log n(x)|$.

In what follows, we assume that $V(x)=V(-x)$ so that in the ground state the gas is confined in a region $x\in[-R,R]\subseteq [-L/2,L/2]$, with $R$ being the ``radius'' of the atomic cloud. The generalization to non-symmetric traps is straightforward.

In the Bethe ansatz description, LDA is implemented through a position-dependent rapidity cutoff $\lambda_F(x)$ in Eqs.~\eqref{eq:rootdensity} and~\eqref{eq:dressing}, fixed such that $n_\text{LDA}(x)=\int_{-\lambda_F(x)}^{\lambda_F(x)} d\lambda \ \rho(\lambda)$. Adopting a grand-canonical description for the gas locally allows one to determine $\lambda_F(x)$ from $V(x)$ via the self-consistent equation $e^{\rm dr}[\pm \lambda_F(x)] = 0$, with the dressed energy $e^{\rm dr}(\lambda)$ satisfying
\be\label{eq:dr_energ}
e^{\rm dr}(\lambda) = \frac{\lambda^2}{2} -\mu + V(x) + \int_{-\lambda_F(x)}^{\lambda_F(x)} \frac{d\lambda'}{2\pi} \frac{2c\  e^{\rm dr}(\lambda')}{c^2+(\lambda-\lambda')^2}.
\ee
In Eq.~\eqref{eq:dr_energ}, the chemical potential $\mu$ is chosen so that the inhomogeneous gas contains exactly $N=\int dx\ n_\text{LDA}(x)$ particles.

Implementing the LDA in the Bethe ansatz framework results in a position-dependent Fermi surface, on top of which low-energy fluctuations can be incorporated like in the standard Luttinger liquid theory reviewed in Sec.~\ref{sec:LLreview}. This leads to the so-called inhomogeneous Luttinger liquid Hamiltonian \cite{Citro2008, GHOSH2006, brun2017one, dubail2017conformal, scopa2021exact, scopa2020one, brun2018inhomogeneous, ruggiero2021quantum, ruggiero2020quantum,Gluza2022, Moosavi2021,tajik2023experimental,moosavi2023exact}
\begin{equation} \label{eq:LL_inhom}
        \hat{\mathscr{H}}_\text{inh} =\int_0^L \frac{dx \ v_\text{LDA}(x) }{2 \pi }  \left( K(x)\left[\partial_x \hat{\theta}(x)\right]^2 + \frac{\left[\partial_x \hat{\phi}(x)\right]^2}{K(x)} \right).
\end{equation}
Here, $K(x)= [1^{\rm dr}(\lambda_F(x)]^2$ is the local Luttinger parameter and $v_\text{LDA}=\pi n_\text{LDA}(x)/K(x)$.

Notice that the Hamiltonian \eqref{eq:LL_inhom} is still quadratic in the fluctuating fields. Thus, exploiting Wick's theorem, it is possible to derive a generic expression for the long-distance asymptotics of the OBDM in the presence of confining potentials
\be\label{eq:main-OBDM-trap}
g_1(x,y)=e^{G_{\theta\theta}(x,y)} \prod_{z=x,y} \frac{\sqrt{B_0(z)}n_\text{LDA}(z)^{\frac{2K(z)-1}{4K(z)}}}{[\Ltilde v_\text{LDA}(z)]^{\frac{1}{4K(z)}}} e^{-\frac{1}{2}G_{\theta\theta}(z)},
\ee
at leading order in the harmonic expansion ($m=0$), and valid for $|x-y|\gg \max[d_\text{LDA}(x),d_\text{LDA}(y)]$, with $d_\text{LDA}(x)=d(n_\text{LDA}(x))$. Here, $B_0(x) = B_0(c/n_\text{LDA}(x))$ is the local nonuniversal amplitude, and $G_{\theta\theta}(x)$ and $G_{\theta\theta}(x,y)$ are the Green's functions of the phase fluctuating field $\hat\theta(x)$. We also introduced the timescale 
\be
\Ltilde=\int_{-R}^R \frac{dx}{v_\text{LDA}(x)}
\ee
associated to the inhomogeneous Luttinger liquid model~\eqref{eq:LL_inhom}, which is the time needed by an excitation with velocity $v_\text{LDA}(x)$ to propagate from one edge to the other of the atomic cloud.

We stress that $\langle\hat\theta(x)\hat\theta(y)\rangle$ is the phase-phase expectation value computed on the inhomogenous equilibrium state (at either zero or finite temperature) of the trapped Lieb-Liniger gas \eqref{eq:LL-model-trap}, while $G_{\theta\theta}(x,y)=\langle\hat\theta[s(x)]\hat\theta[s(y)]\rangle$ is the corresponding correlation after the change of coordinate $x\to s(x)$ that maps the modulated Fermi surface onto one with unit sound velocity and local Luttinger parameter $K[s(x)]$ (see, e.g., Refs.~\cite{scopa2021exact, Scopa_2023, dubail2017conformal} and the discussion below). One then needs to define a regularized Green's function for phase-phase correlations occurring at same position $x$~\cite{brun2018inhomogeneous, scopa2020one}
\be
G_{\theta\theta}(x)=\lim_{x\to x'} \left[G_{\theta\theta}(x,x')-G_{\theta\theta}^\text{hom}(x-x')\right]
\ee
where ultraviolet divergences are removed by exploiting the known result for the homogeneous gas, namely $G_{\theta\theta}^\text{hom}(x)=-K[s(x)]/4 \log|s(x)|^2$~\cite{giamarchi2003quantum}.

Equation~\eqref{eq:main-OBDM-trap} readily provides the long-distance asymptotics of the OBDM in terms of $G_{\theta\theta}(x,y)$. In general, analytical results for $G_{\theta\theta}(x,y)$ are not available, so we treat this function as an input to our theory. In Sec.~\ref{sec:nummethod}, we discuss an efficient numerical implementation to obtain $G_{\theta\theta}(x,y)$ for arbitrary potentials, and at finite interaction strengths and temperature. Conversely, the limits of strong and weak interactions are exactly solvable and are discussed in the following paragraphs.

Lastly, we note that Eq.~\eqref{eq:main-OBDM-trap} has singularities in the limit $x\to y$, which parallel those discussed in Sec.~\ref{sec:LLreview} for the homogeneous case. Therefore, the long-distance asymptotics given in Eq.~\eqref{eq:main-OBDM-trap} must be complemented with a short-distance expansion for $|x-y|\ll \min[d_\text{LDA}(x),d_\text{LDA}(y)]$
\begin{align}\label{eq:short-trap}
&g_1(x,y)=n_\text{LDA}(\zeta)\left(1+C_2(\zeta)\left[n_\text{LDA}(\zeta)|x-y|\right]^2\right. \\[4pt]
&\left.\quad +C_3(\zeta)\left[n_\text{LDA}(\zeta)|x-y|\right]^3+C_4(\zeta)\left[n_\text{LDA}(\zeta)|x-y|\right]^4\right),\nonumber
\end{align}
with $\zeta=(x+y)/2$ and the coefficients $C_q(x)\equiv C_q(c/n_\text{LDA}(x))$ obtained as simple LDA extensions of the results discussed in Sec.~\ref{sec:LLreview}. Combining together the two asymptotic results of Eqs.~\eqref{eq:main-OBDM-trap} and~\eqref{eq:short-trap}, we approximate the OBDM at all distances to determine the momentum distribution of the trapped gas~\eqref{eq:LL-model-trap}.

\subsection{Finite-temperature weakly interacting bosons\\ in a harmonic trap}\label{sec:GP}

In the quasicondensate regime $\gamma \to 0^+$, namely, at weak repulsive interactions $c\to0$ and high density $n(x)=\bar{n}(x)/c$ such that $\bar{n}(x)$ is finite, analytical results can be derived for the OBDM in a harmonic trap $V(x)=\frac{1}{2} \omega^2 x^2$, see also Ref.~\cite{Citro2008}. In this regime, our formalism below is equivalent to Bogoliubov theory, see, e.g.,~Ref.~\cite{CastinMora2003}.

Starting from the equation of state, $\mu=c n$ (see, e.g., Ref.~\cite{pitaevskii2016bose}), using the LDA one finds the local density
\begin{equation}
    n_{\text{LDA}} (x)= \frac{\omega^2 R^2}{2c} \left( 1 - \frac{x^2}{R^2} \right),
\end{equation}
where $R=\sqrt{2\mu}/\omega$. Equivalently, $R$ can be related to the number of particles through
\begin{equation}
    N = \frac{2\omega^2 R^3}{6c}.
\end{equation}
In the quasicondensate regime, Bogoliubov theory predicts a divergent Luttinger parameter given by~\cite{cazalilla2004bosonizing}
\begin{equation}
    K(x) \overset{\gamma\to0^+}{\simeq} \frac{\pi}{\sqrt{\gamma(x)}}  =  \pi \sqrt{\frac{n_{\rm LDA}(x)}{c} },
\end{equation}
while the sound velocity is
\begin{equation}\label{eq:sound-v-bogo}
    v_\text{LDA}(x) \overset{\gamma\to0^+}{\simeq} \sqrt{ c \  n_{\rm LDA} (x) } .
\end{equation}

In the special case of a harmonic potential, the Luttinger liquid Hamiltonian~\eqref{eq:LL_inhom} can be diagonalized using the following mode expansion~\cite{Citro2008, Petrov2004,Gluza2022}
\begin{eqnarray}
    \label{eq:mode-exp-bogo-trap1}
    \hat{\theta}(x) & = & i \omega \sqrt{\frac{c}{2}} \sum_{p>0}  \frac{{\frak a}_p(x/R)}{\sqrt{\varepsilon_p} } (\hat{a}^\dagger_p - \hat{a}^{}_p),  \\[4pt]
     \label{eq:mode-exp-bogo-trap2}
    \hat{\phi}(x) & = & \frac{\pi \omega}{\sqrt{2c}} \sum_{p>0}  \frac{{\frak b}_p(x/R)}{\sqrt{\varepsilon_p} } (\hat{a}^\dagger_p + \hat{a}^{}_p).
\end{eqnarray}
with $[\hat{a}^{}_p,\hat{a}_{q}^\dagger]=\delta_{p,q}$, $[\hat{a}^{}_p,\hat{a}^{}_{q}]=0$, and the phonon dispersion in the trap
\begin{equation}
    \varepsilon_p= \omega \sqrt{\frac{p(p+1)}{2}}.
\end{equation}
Importantly, the mode amplitudes ${\frak a}_p(u)$ and ${\frak b}_p(u)$ entering in Eqs.~\eqref{eq:mode-exp-bogo-trap1} and~\eqref{eq:mode-exp-bogo-trap2} have a known analytical expression in terms of the Legendre polynomials ${\cal L}_p(u)$ \cite{Citro2008, Petrov2004,Gluza2022}
\begin{eqnarray}
    {\frak a}_p(u) &=& \sqrt{p+\frac{1}{2}} {\cal L}_p (u), \\[4pt]
    {\frak b}_p(u) &=& -  \sqrt{ \frac{p+1/2}{2p (p+1)} } (1-u^2) {\cal L}_p' (u)  .
\end{eqnarray}
These functions satisfy the differential equation
\begin{equation}
    \left( \begin{array}{cc}
        0 & \partial_u \\
        - \frac{1-u^2}{2} \partial_u & 0
    \end{array} \right) \left( \begin{array}{c}
        {\frak a}_p \\ {\frak b}_p
    \end{array}  \right) \, = \,  \sqrt{\frac{p (p+1)}{2}} \left( \begin{array}{c}
        {\frak a}_p \\ {\frak b}_p
    \end{array}  \right),
\end{equation}
and are normalized such that
\begin{equation}
    \int_{-1}^1 du\ {\frak a}_p (u) {\frak a}_q(u) = \int_{-1}^1 du\  \frac{2 {\frak b}_p(u) {\frak b}_q(u)}{1-u^2} = \delta_{p,q}  .
\end{equation}

Using Eqs.~\eqref{eq:mode-exp-bogo-trap1} and~\eqref{eq:mode-exp-bogo-trap2}, the Hamiltonian \eqref{eq:LL_inhom} becomes diagonal in the mode operators and reads (up to an additive constant)
\begin{equation}
        \hat{\mathscr{H}}_\text{inh}= \sum_{p>0} \varepsilon_p\  \hat{a}^\dagger_p \hat{a}_p.
\end{equation}
The two-point correlation functions of Luttinger fields computed in the ground state in harmonic traps thus have the following analytical expressions
\begin{align}\label{eq:bogo-2pt-functions1}
 & \langle \hat{\theta}(x) \hat{\theta}(y) \rangle =  \frac{c \omega^2}{2}  \sum_{p>0} \frac{{\frak a}_p (x/R) {\frak a}_p (y/R)}{\varepsilon_p},  \\[4pt]
  \label{eq:bogo-2pt-functions2}
   &\langle \hat{\phi}(x) \hat{\phi}(y) \rangle = \frac{\pi^2\omega^2}{2c}  \sum_{p>0} \frac{{\frak b}_p (x/R) {\frak b}_p (y/R)}{\varepsilon_p},  \\[4pt]
   \label{eq:bogo-2pt-functions3}
     &\langle \hat{\phi}(x) \hat{\theta}(y) \rangle =  i\frac{\pi\omega^2}{2}  \sum_{p>0} \frac{{\frak b}_p (x/R) {\frak a}_p (y/R)}{\varepsilon_p}.     
\end{align}
Notice that the sum entering in Eqs.~\eqref{eq:bogo-2pt-functions1}--\eqref{eq:bogo-2pt-functions3} is fast converging in $p$. These formulas can be generalized straightforwardly to finite temperatures by replacing the ground state expectation values $\left< \hat{a}_p \hat{a}_p^\dagger \right> = 1$ and $\left< \hat{a}_p^\dagger \hat{a}_p \right> = 0$ by the thermal equilibrium ones, $\tr(\hat{\rho}_\text{th}\hat{a}_p\hat{a}^\dagger_p)=1+n_\text{BE}(\varepsilon)$ and $\tr(\hat{\rho}_\text{th}\hat{a}^\dagger_p\hat{a}_p)=n_\text{BE}(\varepsilon)$, for a the density matrix $\rho_\text{th}\propto\exp\big(-\beta\hat{\mathscr{H}}_\text{inh}\big)$. The thermal occupation of a bosonic mode is $n_\text{BE}(\varepsilon)=1/(e^{\varepsilon/T}-1)$. One finds that, at finite temperature, Eqs.~\eqref{eq:bogo-2pt-functions1}-\eqref{eq:bogo-2pt-functions2} become
\begin{align}\label{eq:bogo-2pt-functions1_finT}
 & \langle \hat{\theta}(x) \hat{\theta}(y) \rangle =  \frac{c \omega^2}{2}  \sum_{p>0} \frac{{\frak a}_p (x/R) {\frak a}_p (y/R)}{\varepsilon_p} [1+2n_\text{BE}(\varepsilon_p)];  \\[4pt]
  \label{eq:bogo-2pt-functions2_finT}
   &\langle \hat{\phi}(x) \hat{\phi}(y) \rangle = \frac{\pi^2\omega^2}{2c}  \sum_{p>0} \frac{{\frak b}_p (x/R) {\frak b}_p (y/R)}{\varepsilon_p}[1+2n_\text{BE}(\varepsilon_p)], 
\end{align}
while $\langle\hat\phi(x) \hat{\theta}(y)\rangle$ in Eq.~\eqref{eq:bogo-2pt-functions3} remains unchanged.

Equation~\eqref{eq:bogo-2pt-functions1_finT}, together with the known result for the nonuniversal amplitude $B_0\simeq 1$ for $\gamma\to 0^+$ \cite{brun2018inhomogeneous}, gives direct access to the asymptotic long-distance behavior of the OBDM~\eqref{eq:LL-model-trap} in the quasicondensate regime for a harmonic potential. Explicitly,
\begin{align}\label{eq:OBDM-bogo-final}
&g_1(x,x')=\exp\left(\frac{c\omega^2}{2}\sum_{p>0} {\frak a}_p(x){\frak a}_p(x')\frac{[1+2n_\text{BE}(\varepsilon_p)]}{\varepsilon_p}\right)\nonumber \\[4pt]
&\qquad\times\prod_{z=x,x'}\frac{n_\text{LDA}(z)^{\frac{2K(z)-1}{4K(z)}} }{\exp\left(\frac{c\omega^2}{4}\sum_{p>0} {\frak a}^2_p(z)\frac{[1+2n_\text{BE}(\varepsilon_p)]}{\varepsilon_p}\right)}.
\end{align}

%%%%%%%%%%%%%%%%%%%%%%%%%%%%%%%%%%%%%%
\begin{figure}[!t]
  \includegraphics[width=0.98\columnwidth]{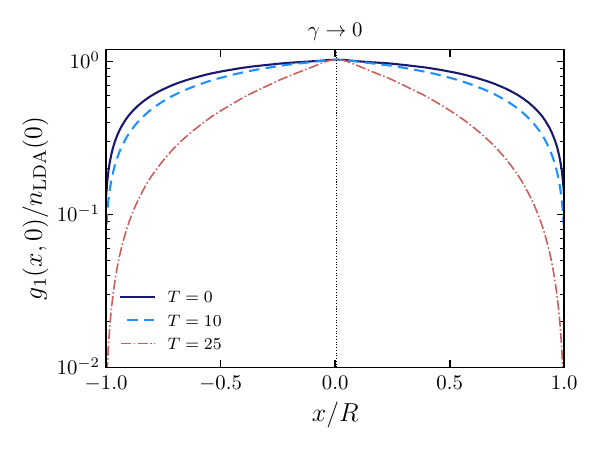}
  \vspace{-0.4cm}
  \caption{Long-distance asymptotics of $g_1(x,x')$ for bosons in the quasicondensate regime ($\gamma\to0$) confined in a harmonic trap $V(x)=\frac{1}{2}\omega x^2$, obtained from Eq.~\eqref{eq:OBDM-bogo-final}. We set the trap frequency to $\omega=1$ and take $N=20$. We plot $g_1(x,x')$ vs $x/R$ ($R$ is the atom cloud radius) for $x'=0$ and for different temperatures (see legend). The temperatures are to be compared with the energy scale $\Ltilde^{-1}\simeq 0.47$ of the inhomogeneous Luttinger liquid model.}
  \label{fig:g1_finT_weak}
\end{figure}
%%%%%%%%%%%%%%%%%%%%%%%%%%%%%%%%%%%%%%
In Fig.~\ref{fig:g1_finT_weak}, we plot the results obtained evaluating Eq.~\eqref{eq:OBDM-bogo-final} at the center of the trap for different temperatures. The same results can be obtained from Eq.~\eqref{eq:main-OBDM-trap} by inserting $G_{\theta\theta}(x,y)$ given in Eq.~\eqref{eq:bogo-2pt-functions1_finT} (upon using the change of coordinates (\ref{eq:stretched-coo}) specified below).

\subsection{Finite-temperature hard-core bosons in a trap}\label{sec:TG-gas-analytical}

In the hard-core boson limit ($\gamma\to\infty$, also known as the Tonks-Girardeau limit~\cite{girardeau1960relationship}), analytical results can be derived for the OBDM of the inhomogeneous gas. The zero-temperature case was discussed in Refs.~\cite{brun2017one, scopa2020one, Scopa_2023}, so here we focus on the finite temperature $T= 1/\beta$ regime. The key feature of the Tonks-Girardeau limit that makes possible an analytical treatment is that the Luttinger parameter is constant and equal to its free-fermionic value, $K(\infty)=1$, regardless of the local value of density. This allows us to reduce the inhomogeneous Luttinger Hamiltonian \eqref{eq:LL_inhom} to the 2D conformal field theory of a compact free boson. In terms of density fluctuating fields, the corresponding action reads \cite{giamarchi2003quantum,cazalilla2004bosonizing,DiFrancesco:1997nk}
\be
\label{eq:action}
\mathscr{S}=\frac{1}{2\pi}\int_0^{\beta/\Ltilde} d\tau \int_0^1 ds \left( \nabla \phi \right)^2,
\ee
where $\nabla= (\partial_s, \partial_\tau)$ is the gradient in 2D Euclidean spacetime. Here $\tau$ is the dimensionless (i.e., rescaled by $1/\Ltilde$) imaginary time, and the field $\phi$ is periodic in that imaginary time direction, $\phi(x, \tau + \beta/\Ltilde ) =  \phi(x, \tau)$. The spatial coordinate $x$ has been replaced with the stretched coordinate~\cite{brun2017one,brun2018inhomogeneous,scopa2020one}
\be\label{eq:stretched-coo}
s(x)=\Ltilde^{-1}\int_{-R}^x \frac{dx'}{v_\text{LDA}(x')},
\ee
which can be interpreted as the (dimensionless) time needed for an excitation traveling from the left boundary to the point $x$. The field $\phi$ has Dirichlet boundary conditions at the two boundaries, $\phi(0,\tau) = \phi(1 , \tau)$. The action (\ref{eq:action}) is then the one of a free boson living on a cylinder of width $1$ and circumference $\beta/\Ltilde$. After some technical manipulations detailed in Appendix~\ref{app:cft-TG-details}, one obtains the following result for the finite-temperature Green's function of the phase fields
 \begin{equation}
G_{\theta\theta}(x,y)= - \frac{1}{2}\log{ \left| \frac{\vartheta_1 \!\left( \frac{s(x) - s(y)}{2} \left|\frac{i\beta}{2\mathcal{T}} \!\right. \right) \vartheta_1 \!\left( \frac{s(x) + s(y)}{2} \left| \frac{i\beta}{2\mathcal{T}} \!\right.\right)}{\left[\partial_z \vartheta_1 \!\left( 0 \left|\frac{i\beta}{2\mathcal{T}} \!\right. \right) \right]^2 } \right| } 
\label{eq:theta_corr_annulus}
\end{equation}
where
\be\label{eq:jacobi-theta}
\vartheta_1(z|\tau)= -i \sum_{r \in {\Bbb Z} + \frac{1}{2}} (-1)^{r - \frac{1}{2}} e^{2\pi r i z } e^{2\pi \frac{r^2}{2} i \tau }
\ee
is the Jacobi $\theta$ function. The regularized Green's function at equal points is then
\begin{eqnarray}
 \nonumber   G_{\theta \theta}(x) &=& \lim_{x' \rightarrow x} \left[ G_{\theta \theta} (x, x') + \frac{1}{4} \log | s(x)-s(x') |^2 \right] \\
    &=& - \frac{1}{2}\log{ \left|  \frac{ \frac{\pi}{2} \vartheta_1\!\left(s(x) \left|\frac{i\beta}{2\mathcal{T}}\!\right.\right)}{\partial_z \vartheta_1\!\left( 0 \left|\frac{i\beta}{2\mathcal{T}}\!\right.\right) } \right| } .
    \label{eq:reg_green_analytics}
\end{eqnarray}
%%%%%%%%%%%%%%%%%%%%%%%%%%%%%%%%%%%%%%
\begin{figure}[!t]
  \includegraphics[width=0.98\columnwidth]{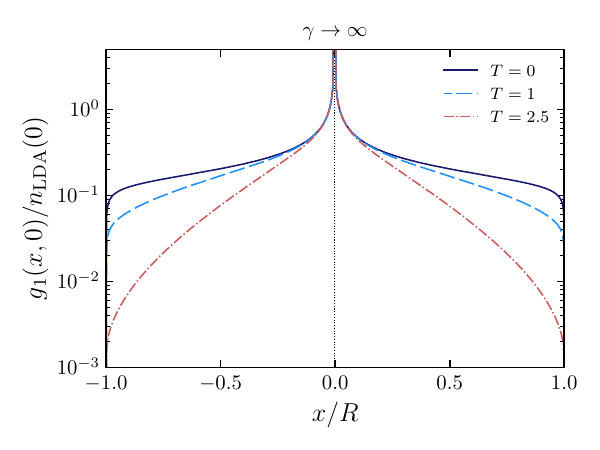}
  \vspace{-0.4cm}
  \caption{Long-distance asymptotics of $g_1(x,x')$ for hard-core bosons ($\gamma\to\infty$) confined in a harmonic trap $V(x)=\frac{1}{2}\omega x^2-\mu$, obtained from Eq.~\eqref{eq:main-OBDM-trap} and \eqref{eq:theta_corr_annulus}. We set the trap's frequency $\omega=1$ and fix $\mu$ such that $N=20$. We plot $g_1(x,x')$ vs $x/R$ ($R=\sqrt{2\mu/\omega}$ is the atom cloud radius) for $x'=0$ and for different temperatures (see legend). The temperatures are to be compared with the energy scale $\Ltilde^{-1}\simeq 0.2$ of the inhomogeneous Luttinger liquid model.}
  \label{fig:g1_finT_TG}
\end{figure}
%%%%%%%%%%%%%%%%%%%%%%%%%%%%%%%%%%%%%%
By inserting Eqs.~\eqref{eq:theta_corr_annulus} and~\eqref{eq:reg_green_analytics} in Eq.~\eqref{eq:main-OBDM-trap}, and using that the nonuniversal amplitude $B_0(\infty)=\text{G}(3/2)^4/\sqrt{2\pi}\simeq 0.5214$ in the Tonks-Girardeau limit, where $\text{G}(\cdot)$ is the Barnes G-function~\cite{lenard1964momentum, lenard1972some, forrester2003finite, widom1973toeplitz}, one finds the long-distance asymptotics of the trapped gas~\eqref{eq:LL-model-trap},
\begin{align}\label{eq:g1-tg-thermal}
 &g_1(x,y) =  \frac{B_0(\infty)}{\sqrt{2\Ltilde}}  \\ & \quad \times \frac{ \left|\partial_z \vartheta_1 \!\left( 0 \left|\frac{i\beta}{2\mathcal{T}} \!\right. \right)\right|^\frac{1}{2} \left| \vartheta_1 \!\left( s(x) \left|\frac{i\beta}{2\mathcal{T}} \!\right. \right) \right|^\frac{1}{4} \left| \vartheta_1 \!\left( s(y) \left|\frac{i\beta}{2\mathcal{T}} \!\right. \right) \right|^\frac{1}{4} }{\left| \vartheta_1 \!\left( \frac{s(x) - s(y)}{2} \left| \frac{i\beta}{2\mathcal{T}} \!\right. \right) \right|^\frac{1}{2} \left| \vartheta_1 \!\left( \frac{s(x) + s(y)}{2} \left| \frac{i\beta}{2\mathcal{T}} \!\right. \right) \right|^\frac{1}{2}}.\nonumber
\end{align}
We plot this result in Fig.~\ref{fig:g1_finT_TG} for a harmonic potential. In the zero-temperature limit $\beta \to \infty$, Eq.~\eqref{eq:g1-tg-thermal} reduces to the expression in Ref.~\cite{forrester2003finite,brun2017one}.

For completeness, we also report the expression of the Green's function for the density fluctuating fields, $G_{\phi\phi}(x,y)\equiv\langle\hat\phi[s(x)]\hat\phi[s(y)]\rangle$, entering, e.g., in the calculation of the density ripples and in the density-density correlations, see for instance Refs.~\cite{ruggiero2021quantum, urilyon2024quantum}, or   Appendix~\ref{app:cft-TG-details} for a derivation:
\begin{equation}
 G_{\phi\phi}(x,y)= - \frac{1}{2}\log{ \left| \frac{\vartheta_1 \!\left( \frac{{ s(x) - s(y)}}{2} \left|\frac{i\beta}{2\mathcal{T}} \!\right. \right)}{\vartheta_1 \!\left( \frac{{ s(x) + s(y)}}{2} \left| \frac{i\beta}{2\mathcal{T}} \!\right. \right)}  \right| }.
    \label{eq:phi_corr_annulus}
\end{equation}

\section{Finite-temperature Green's functions of the inhomogeneous Luttinger liquid}
\label{sec:nummethod}

In this section we discuss the numerical method used to determine the equilibrium two-point correlation functions of the Luttinger fields $\hat\theta(x)$ and $\hat\phi(x)$ in the Hamiltonian~\eqref{eq:LL_inhom}, valid for arbitrary strengths of the contact interaction at finite temperature. We follow Ref.~\cite{Bastianello_2020}, where a numerical method is described for the zero-temperature Green's functions (see also Ref.~\cite{scopa2020one}), and extend that method to finite temperature.

Our starting point is the Hamiltonian \eqref{eq:LL_inhom}, which for convenience we express in terms of the stretched coordinate $s(x)$ in Eq.~\eqref{eq:stretched-coo}, with $v_\text{LDA}(x)=\pi n_\text{LDA}(x)/K(x)$,
\be
\hat{\mathscr{H}}_\text{inh} =\frac{1}{2 \pi\Ltilde }\int_0^1 ds \left(\pi^2 K(s)\hat\Pi^2(s) + \frac{\left[ \partial_s \hat{\phi}(s) \right]^2}{K(s)} \right).
\ee
Here, we introduced the canonically conjugated momentum $\hat\Pi(s)=\de_s\hat\theta(s)/\pi$ such that $[\hat\Pi(s),\hat\phi(s)]=-i\delta(s-s')$. This Hamiltonian can be readily discretized as follows
\be\label{eq:discr-LL}
\hat{\mathscr{H}}_\text{inh} =\frac{\pi}{2M\Ltilde} \sum_{j=1}^M K_j \hat\Pi_j^2 +\frac{M}{\pi\Ltilde}\sum_{j=1}^{M+1} \frac{(\hat\phi_j-\hat\phi_{j-1})^2}{K_j+K_{j-1}},
\ee
where $M\gg 1$ is the number of sampling points in the unit interval, and $K_j\equiv K(s_j)$ is the discretized Luttinger parameter. The Luttinger fields are replaced with their discretized version $\hat\Pi_j$, $\hat\phi_j$ satisfying $[\hat\Pi_j,\hat\phi_{j'}]=-i\delta_{j,j'}$. Open boundary conditions are imposed on the chain, implying that $\hat\phi_0=\hat\phi_{M+1}=0$ and $K_0=K_{M+1}=1$. In matrix form,
\be
\hat{\mathscr{H}}_\text{inh} =\hat{\bm\Phi}^\text{T} \ h \ \hat{\bm\Phi},
\ee
where
\be\label{eq:lutt-field-mat}
\bm\Phi^\text{T} = \begin{pmatrix} \hat\phi_1 , \dots,   \hat\phi_M , \hat\Pi_1 , \dots , \hat\Pi_M\end{pmatrix},
\ee
and $h$ is the $2M\times 2M$ Hamiltonian matrix having nonvanishing elements
\begin{align}
&h_{i,j}=\frac{M}{\pi\Ltilde}\left[\frac{\delta_{i,j}-\delta_{i,j+1}-\delta_{i+1,j}}{K_j+K_{j-1}} +\frac{\delta_{i,j}}{K_{j+1}+K_{j}}\right];\\[4pt]
&h_{i+M,j+M}=\frac{\pi}{2M\Ltilde} \delta_{i,j} K_j,
\end{align}
for $i,j=1,\dots, M$. It is convenient to change the operator basis from $\hat{\bm\Phi}$ to the bosonic modes
\be
\hat{b}^\pm_j=\frac{\hat\phi_j\pm i \hat\Pi_j}{\sqrt{2}},
\ee
satisfying $[\hat{b}^+_j, \hat{b}^-_{j'}] = \delta_{j,j'}$ and commuting otherwise, that we collect in the $2M$-vector
\be
\hat{\bm{b}}^\dagger\equiv\begin{pmatrix} \hat{b}^-_1 ,  \dots , \hat{b}^-_M , \hat{b}^+_1 , \dots , \hat{b}^+_M\end{pmatrix}=\hat{\bm\Phi}^\text{T} W^\dagger
\ee
with $2M\times 2M$ matrix $W$ having nonvanishing elements for $j=1,\dots,M$
\begin{align}
&W_{j,j}\equiv W_{j,j+M}=1/\sqrt{2},\nonumber\\[4pt]
&W_{j+M,j}=i/\sqrt{2}, \quad W_{j+M,j+M}=-i/\sqrt{2}.
\end{align}

This notation allows us to recast the Hamiltonian~\eqref{eq:discr-LL} in the quadratic form
\be\label{eq:LL-discr2}
\hat{\mathscr{H}}_\text{inh} =\hat{\bm{b}}^\dagger  \left(W^\dagger \ h\ W\right)  \hat{\bm{b}},
\ee
which can be diagonalized by a further unitary transformation $U$. Denoting $\hat{\bm\eta}=U\hat{\bm b}$, one has
\begin{align}
\hat{\mathscr{H}}_\text{inh} &=\hat{\bm\eta}^\dagger U \left(W^\dagger \ h\ W\right) U^\dagger \hat{\bm\eta}=\sum_{j=1}^{M} \varepsilon_j (\hat{\eta}^\dagger_j \hat{\eta}_j + \hat{\eta}_j \hat{\eta}_j^\dagger),
\end{align}
with $\hat{\bm\eta}^\dagger=(\hat\eta^\dagger_1,\dots,\hat\eta^\dagger_M,\hat\eta_1,\dots,\hat\eta_M)$ and eigenvalues $\varepsilon_j\equiv \varepsilon_{j+M}$ for $j=1,\dots, M$, following from the symplectic structure of $U$ required to preserve the canonical commutation relations of $\hat{b}^\pm_j$ operators \cite{Bastianello_2020}. 

Given the structure of the Fock space, it is convenient to consider the associated matrix
\be
\tilde{h}=\begin{pmatrix} \text{Id} & 0 \\ 0 & -\text{Id} \end{pmatrix}\left(W^\dagger \ h\ W\right),
\ee
where $\text{Id}$ is the $M\times M$ identity matrix and projects on the negative eigenvalue spectrum of $\tilde{h}$ such that the degeneracy of the spectrum is removed.

Denoting as $\gamma_j$ the eigenvectors, and as $\omega_j<0$ the eigenvalues, of $\tilde{h}$ restricted to the negative energy subspace, the two-point correlation matrix can be written as
\be\label{eq:num-corr}
\langle \hat{b}^+_i \hat{b}^-_j \rangle = \gamma^\dagger_i \ [{\cal P}_T(\omega)]_{ij} \ \gamma_j,
\ee
where ${\cal P}(T)$ is a $M\times M$ diagonal matrix that projects onto the target state over which the expectation value is computed. In a thermal state, it gives Bose-Einstein weights to the bosonic modes
\begin{align}
&[{\cal P}_T(\omega)]_{i,j} = \frac{\delta_{i,j}}{1-e^{-\omega_j/T}}.
\end{align}

Equation~\eqref{eq:num-corr} fixes the structure of the desired correlation function up to a normalization of the eigenvectors $\gamma_j$. In order to fix it, we consider the auxiliary (Hermitian) matrix
\be
O_{ij}= \gamma_i^\dagger \begin{pmatrix} \text{Id} & 0 \\ 0 & -\text{Id} \end{pmatrix} \gamma_i
\ee
with eigenvectors $u_j$, and we construct the normalized eigenvectors as \cite{Bastianello_2020,scopa2020one}
\be
v_j=\frac{u_j}{\sqrt{u_j^\dagger\begin{pmatrix} -\text{Id} & 0 \\ 0 & \text{Id} \end{pmatrix} u_j }}, \quad j=1,\dots,M.
\ee
Finally, we consider the $2M\times 2M$ canonical transformation implemented by 
\be
\Omega=\left(\begin{array}{ccc|}
v^\dagger_1(M+1) & \dots & v^\dagger_1(2M) \\
\vdots & \vdots & \vdots  \\
v^\dagger_M(M+1) & \dots & v^\dagger_M(2M)\\[4pt]
v^\dagger_1(1) & \dots & v^\dagger_1(M) \\
\vdots & \vdots & \vdots\\
v^\dagger_M(1) & \dots & v^\dagger_M(M) \\
\end{array}
\begin{array}{ccc}
v_1 & \dots & v_M
\end{array}
\right),
\ee
in terms of which
\be
\Omega^\dagger \ \tilde{h}\ \Omega= \begin{pmatrix} \varepsilon_1 \\ & \ddots \\ && \varepsilon_M \\[4pt] &&& \varepsilon_1 \\ &&&& \ddots \\ &&&&& \varepsilon_M \end{pmatrix}.
\ee

By returning to the Luttinger fields~\eqref{eq:lutt-field-mat}, we are now in the position of obtaining the desired correlation simply as
\be
\langle \hat{\bm\Phi}^\text{T} \hat{\bm\Phi}\rangle =W \ \Omega\begin{pmatrix}\text{Id}+{\cal P}_T(2\varepsilon)&  \\   & {\cal P}_T(2\varepsilon) \end{pmatrix} \Omega^\dagger \ W^\dagger,
\ee
with the factor $2$ in the thermal state projectors following by our choice of normalization. Explicitly, this matrix contains the Green's functions
\be
\langle \hat{\bm\Phi}^\text{T} \hat{\bm\Phi}\rangle=\begin{pmatrix}
\left[\langle \hat\phi_i \hat\phi_j\rangle\right]_{i,j=1}^M &  \left[\langle \hat\phi_i \hat\Pi_j\rangle\right]_{i,j=1}^M  \\[8pt]
\left[\langle \hat\Pi_i \hat\phi_j\rangle\right]_{i,j=1}^M  & \left[\langle \hat\Pi_i \hat\Pi_j\rangle\right]_{i,j=1}^M 
\end{pmatrix},
\ee
so we perform the numerical integration 
\be
{\cal C}= A \ \langle \hat{\bm\Phi}^\text{T} \hat{\bm\Phi}\rangle \ A^\text{T},
\ee
with
\be
A=\left(\begin{array}{c|ccccc}
\text{Id} \\
\hline 
& 1\\
&\frac{\pi}{M} & 1\\
&\frac{\pi}{M}& \frac{\pi}{M} & 1\\
&\vdots& &\ddots& \ddots\\
&\frac{\pi}{M}&\dots&&\frac{\pi}{M}& 1
\end{array}\right),
\ee
yielding the Green's functions for density and phase fluctuating fields, $x_j\equiv x(s_j)$,
\be\label{eq:corr-matrix-final}
{\cal C}=\begin{pmatrix}
\left[G_{\phi\phi}(x_i,x_j)\right]_{i,j=1}^M &  \left[G_{\phi\theta}(x_i,x_j)\right]_{i,j=1}^M  \\[8pt]
\left[G_{\theta\phi}(x_i,x_j)\right]_{i,j=1}^M  & \left[G_{\theta\theta}(x_i,x_j)\right]_{i,j=1}^M 
\end{pmatrix}.
\ee

Lastly, the regularized Green's functions are obtained from those in Eq.~\eqref{eq:corr-matrix-final} by removing ultraviolet divergences affecting the correlation at coincindent points. Focusing on $G_{\theta\theta}(x,x')$:
\be\label{eq:G-thth-reg1}
G_{\theta\theta}(x_j)=G_{\theta\theta}(x_{j+1},x_{j-1})+\frac{K_j}{4}\log\left( \frac{2}{2+M}\right)^2
\ee
for $j\neq 1, M$, and at the boundaries
\begin{eqnarray}\label{eq:G-thth-reg2}
 G_{\theta\theta}(x_1)&=&G_{\theta\theta}(x_1,x_2)+\frac{K_2}{4}\log\!\left(\frac{1}{2+M}\right)^{2}\!,\\
 G_{\theta\theta}(x_M)&=&G_{\theta\theta}(x_M,x_{M-1})+\frac{K_{M+1}}{4}\log\!\left(\frac{1}{2+M}\right)^2\!\! .\quad\ \ \ \label{eq:G-thth-reg3}
\end{eqnarray}
The regularization of the other Green's function $G_{\phi \phi}$ is done in a similar way.

The phase-phase Green's function in Eq.~\eqref{eq:corr-matrix-final} [together with its regularization in Eqs.~\eqref{eq:G-thth-reg1}--\eqref{eq:G-thth-reg3}] is used as a numerical input for Eq.~\eqref{eq:main-OBDM-trap} to determine the OBDM~\eqref{eq:LL-model-trap} in arbitrary potentials, and for finite temperature and interactions. 

\begin{figure}[!t]
\includegraphics[width=0.98\columnwidth]{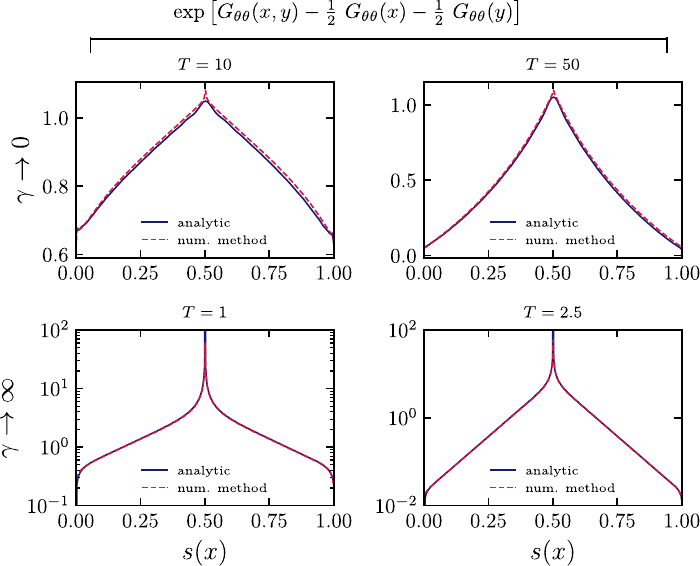}
\vspace{-0.1cm}
\caption{Results for $\exp[G_{\theta\theta}(x,y)-\frac{1}{2}G_{\theta\theta}(x)-\frac{1}{2}G_{\theta\theta}(y)]$ [entering in the general formula for the OBDM in Eq.~\eqref{eq:main-OBDM-trap}] for $s(y)=0.5$ obtained numerically (dashed lines) and analytically (solid lines) in the limiting cases of $\gamma\to 0 $ (top panels, see Sec.~\ref{sec:GP}) and $\gamma\to\infty$ (bottom panels, see Sec.~\ref{sec:TG-gas-analytical}). The temperatures for which the results are reported are to be compared with the energy scale $\Ltilde^{-1}$ of the inhomogeneous Luttinger liquid model reported in Fig.~\ref{fig:g1_finT_TG} (Fig.~\ref{fig:g1_finT_weak}) for $\gamma \to \infty$ ($\gamma \to 0$) respectively, and increases from left to right.}
\label{fig:greensfunc}
\end{figure}

In Fig.~\ref{fig:greensfunc}, we compare the numerical Green's functions with the analytical results available in the limiting cases $\gamma\to 0$ and $\gamma\to\infty$, respectively. One can see that there is an excellent agreement for the temperatures shown. In the following section, we benchmark the validity of our approach for finite interaction strengths and different temperatures against quantum Monte Carlo simulations.

%%%%%%%%%%%%%%%%%%%%%%%%%%%
%%%%%%%%%%%%%%%%%%%%%%%%%%%

%@@@@@@@@@@@@@@@@@@@@@@@@@@@@@@@@@@@@@@
\begin{figure*}[!t]
\includegraphics[width=\textwidth]{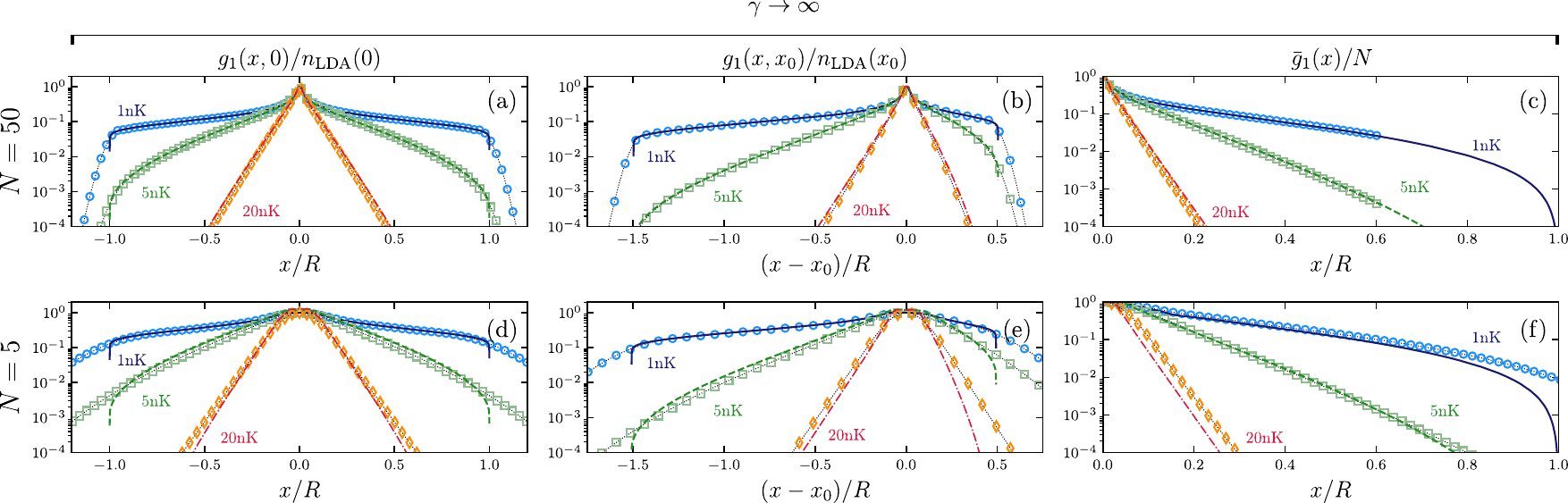}
  \vspace{-0.5cm}
\caption{OBDM of a gas of $^{162}$Dy atoms in the hard-core regime $\gamma\to\infty$ confined in a harmonic trap with frequency $36.4$Hz. We show results for $g_1$ at [(a) and (d)] the trap's center ($x_0=0$), [(b) and (e)] $x_0=R/2$, and [(c) and (f)] the average $g_1$ [see Eq.~\eqref{eq:avg-OBDM}]. Panels (a)--(c) [(d)--(f)] show results for $N=50$ ($N=5$) bosons. In each panel, symbols joined by thin dotted lines are used for the numerical results obtained in the low-density lattice calculations (referred to in the text as the ``{\it exact results}'') and thick lines are used for the minimum between the result from Eq.~\eqref{eq:g1-tg-thermal} (the long-distance asymptotics) and the result from Eq.~\eqref{eq:short-trap} (the short-distance expansion), which is referred to in the text as the results within ``{\it our approach}''. We report results for three temperatures: $T=1,\,5,$ and 20 nK.}\label{fig:hcb-g1}
\end{figure*}
%@@@@@@@@@@@@@@@@@@@@@@@@@@@@@@@@@@@@@@

\section{Finite-temperature OBDM and momentum distribution of trapped bosons}\label{sec:numerical}

Next, we benchmark our results for the OBDM and the momentum distribution. We restore the physical values of the fundamental constants and use the parameters associated to recent experiments to test our results. Specifically, we consider a gas of $^{162}$Dy atoms confined in a harmonic trap with a frequency $\omega= 2\pi \times 36.4$ Hz~\cite{li2023dipolar}.

We first consider the hard-core ($\gamma\to\infty$) limit, for which we compare the results obtained using the analytical approach discussed in Sec.~\ref{sec:TG-gas-analytical} to numerical results obtained in the low-density limit of lattice hard-core boson calculations~\cite{rigol_05}. The latter approach was used to describe experimental results in- and out-of-equilibrium in Refs.~\cite{li2023dipolar, yang2023phantom}. Next, we consider the soft-core [$\gamma\sim {\cal O}(1)$] case, for which we compare the numerical results obtained using the approach discussed in Sec.~\ref{sec:nummethod} to those of quantum Monte Carlo simulations. For convenience in the discussions in this section, we refer to the approaches discussed in Secs.~\ref{sec:TG-gas-analytical} and~\ref{sec:nummethod} as ``{\it our approach}'' and to the unbiased numerical calculations as the ``{\it exact results}''.

\subsection{Hard-core bosons}\label{sec:LLvsExactHCB}

In Fig.~\ref{fig:hcb-g1} we report the results for the OBDM in the hard-core ($\gamma\to\infty$) limit. The results reported for {\it our approach} are those of the regularized OBDM, namely, the minimum between the result from Eq.~\eqref{eq:g1-tg-thermal} (the long-distance asymptotics) and the result from Eq.~\eqref{eq:short-trap} (the short-distance expansion).

In Figs.~\ref{fig:hcb-g1}(a)--\ref{fig:hcb-g1}(c), we show results obtained for $N=50$ hard-core bosons at three temperatures ($T=1,\,5,$ and 20 nK) computed with respect to two different positions in the trap [at the trap center ($x_0=0$) in Fig.~\ref{fig:hcb-g1}(a) and at $x_0=R/2$ in Fig.~\ref{fig:hcb-g1}(b)] as well as the average
\be\label{eq:avg-OBDM}
\bar{g}_1(x)\equiv \frac1R \int dx_0 \ g_1(x_0,x_0-x),
\ee
which is shown in Fig.~\ref{fig:hcb-g1}(c). The agreement between {\it our approach} (continuous lines) and the {\it exact results} (symbols) is excellent at the lowest temperatures shown ($T=1$ and 5 nK) and, as expected, worsens with increasing the temperature. For $T=20$ nK, while the differences are still small, they become visible in the plots.  

The fact that {\it our approach} describes the {\it exact results} for $N=50$ in Figs.~\ref{fig:hcb-g1}(a)--\ref{fig:hcb-g1}(c) so well is remarkable given that its use is justified only in the limit of large number of particles ($N\gg 1$). In experiments with ultracold gases in 2D optical lattices, such as those in Refs.~\cite{wilson2020observation, malvania2021generalized, guo20231d2dcross, li2023dipolar, le2023observation, guo2024cooling, yang2023phantom}, arrays with thousands of 1D gases are created with different number of atoms across them (the population of the 1D gases is maximal at the center of the arrays). For example, for the ones involving $^{162}$Dy atoms in Refs.~\cite{li2023dipolar, yang2023phantom}, the central 1D gases had up to about 40 atoms. The benchmark in Figs.~\ref{fig:hcb-g1}(a)--\ref{fig:hcb-g1}(c) shows that {\it our approach} is suited to describe the OBDM of 1D gases with $\sim$50 atoms for experimentally relevant temperatures.

To explore the effect of decreasing the number of particles to a few atoms, in Figs.~\ref{fig:hcb-g1}(d)--\ref{fig:hcb-g1}(f) we show results for $N=5$. Even for such a small number of particles {\it our approach} is a good approximation for the {\it exact results}, specially at the center of the trap [Fig.~\ref{fig:hcb-g1}(d)] and for $x_0=R/2$ [Fig.~\ref{fig:hcb-g1}(e)]. The deviations for the average $\bar{g}_1(x)$ [Fig.~\ref{fig:hcb-g1}(f)] are larger because, due to finite-size effects, the extent of the cloud is larger in the {\it exact results}. Notice the increase in the support of the correlations from the top ($N=50$) to the bottom ($N=5$) panels in the {\it exact results}.

\begin{figure}[!t]
    \includegraphics[width=\columnwidth]{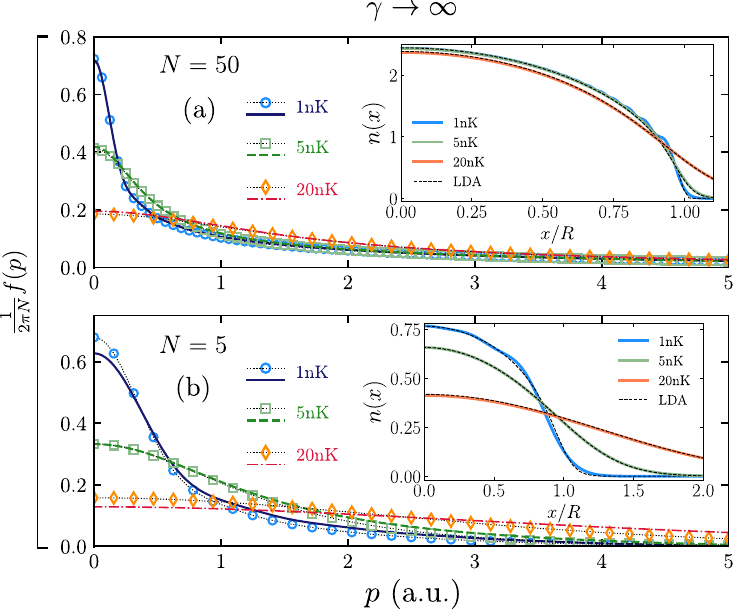}
    \vspace{-0.5cm}
    \caption{(Main panels) Momentum distribution for the same setup of Fig.~\ref{fig:hcb-g1}, obtained via a Fourier transform of the OBDM. In (a) we show the results for $N=50$ and in (b) for $N=5$. Symbols joined by thin dotted lines show the numerical results obtained in the low-density lattice calculations, and thick lines show the results by our approach. (Insets) Same as the main panels but for the density as a function of $x/R$, where $R$ is the atomic cloud radius. Exact numerical results are shown by thick lines, and the LDA approximation by thin dashed lines. We report results for three temperatures: $T=1,\,5,$ and 20 nK.}
    \label{fig:hcb-md}
\end{figure}

In Fig.~\ref{fig:hcb-md} we report the corresponding results for the momentum (main panels) and density (insets) distributions in the hard-core ($\gamma\to\infty$) limit. The momentum distribution calculations within {\it our approach} are carried out computing the Fourier transform of the regularized OBDM. The differences between {\it our approach} and the {\it exact results} are small and difficult to see in the plots for $N=50$ [see Fig.~\ref{fig:hcb-md}(a)], and remain small but become visible for $N=5$ [see Fig.~\ref{fig:hcb-md}(b)].

%@@@@@@@@@@@@@@@@@@@@@@@@@@@@@@@@@@@@@@
%@@@@@@@@@@@@@@@@@@@@@@@@@@@@@@@@@@@@@@

For the integrated difference
\begin{equation}\label{eq:errorD}
    \Delta (p_\text{max}) = \frac{\int_{-p_\text{max}}^{\ p_\text{max}} dp\, |f_\text{exact}(p)-f_\text{appr.}(p)|}{\int_{-p_\text{max}}^{\ p_\text{max}} dp\ f_\text{exact}(p)}
\end{equation}
between the momentum distribution obtained within {\it our approach} ($f_\text{appr.}$) and the {\it exact results} ($f_\text{exact}$), we find the results reported in Table~\ref{tab:tg}. For $N=50$, the differences $\Delta$ are smaller than the differences between the experimental and the model results in Ref.~\cite{li2023dipolar}. For $N=5$ and $T=1,$ 5 nK ($T=20$ nK), they are slightly (significantly) larger than the differences between the experimental and the model results in Ref.~\cite{li2023dipolar}. 

%@@@@@@@@@@@@@@@@@@@@@@@@@@@@@@@@@@@@@@
\begin{table}[!h]
\centering
\begin{tblr}{hlines,hline{1,Z} = 1pt,hline{2,3}=1pt,colspec={|r|ccc|}}
   \SetCell[c=4]{c}          $\Delta$ for hard-core bosons  \\
   \hline
$T$ (nK) &1 &  5 & 20 \\
$N=50$   & 0.033 & 0.033 & 0.039     \\
$N=5$   & 0.138 & 0.102 & 0.253   \\
\end{tblr}
    \caption{Integrated difference $\Delta(p_{\max})$ in Eq.~\eqref{eq:errorD} for hard-core bosons for different temperatures (different columns) and different number of particles (different rows). We set $p_{\max}=8$. }
    \label{tab:tg}
\end{table}
%%%%%%%%%%%%%%%%%%%%%%%%%%%%%%%%%%%%%%%%%%%%

\begin{figure}[!t]
    \centering
    \includegraphics[width=.85\columnwidth]{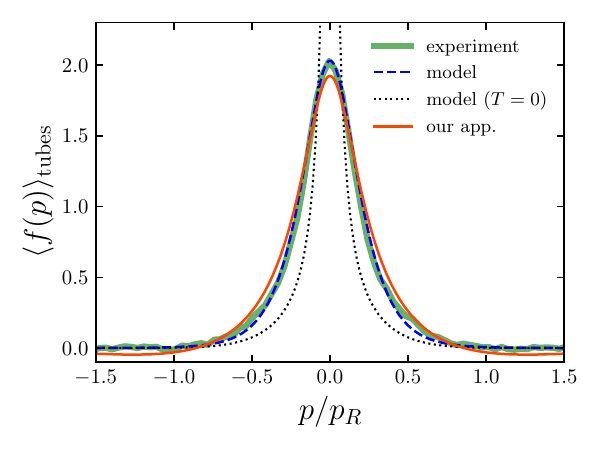}
    \vspace{-0.4cm}
    \caption{Averaged momentum distribution over the array of 1D dysprosium gases in the hard-core regime considered in Ref.~\cite{li2023dipolar}, in which the contact interaction strength is $c = 263\,  \mu{\rm m}^{-1}$ and the harmonic trap frequency is $36.4$ Hz. The experimental data (thick solid line) and model results from the same {\it exact approach} used here (dashed line), reported in Fig.~3(a) of Ref.~\cite{li2023dipolar}, are compared to our approach (thin solid line). To highlight the role of temperature in the experiment, we also plot the model predictions for the same array of 1D gases in the ground state (thin dotted line). The average momentum distribution is computed as $\langle f(p)\rangle_\text{tubes} = \frac{1}{{\cal N}} \sum_N w_N f_{[N,T_N]}(p)$, where $w_N$ is the number of tubes containing $N$ atoms, and $f_{[N,T_N]}(p)$ is the momentum distribution for a fixed number of atoms $N$ and temperature $T_N$. The normalization factor ${\cal N}$ ensures that $\int \langle f(p)\rangle_\text{tubes}  dp = 1$. Momentum is rescaled by the recoil momentum $p_R = 2\pi/741$ nm. See Ref.~\cite{li2023dipolar} for more information about the experimental setup.}
    \label{fig:compare_exp_hard}
\end{figure}

In Fig.~\ref{fig:compare_exp_hard}, we compare experimental measurements of the momentum distribution in the Tonks-Girardeau regime and their modeling based on the {\it exact approach} used in this section, which were reported in Fig.~3(a) of Ref.~\cite{li2023dipolar}, to the momentum distribution obtained using {\it our approach}. We compute the average over 1D gases with the same number of particles and temperatures as in the modeling discussed in detail in Ref.~\cite{li2023dipolar}. We find that there is a good agreement despite the fact that in the experiments $\approx$33\% of the particles are 1D gases with 5 or fewer particles, and that the average temperature of the gases is $\approx$10 nK. As anticipated in Sec.~\ref{sec:LLreview}, even though Eq.~\eqref{eq:short-trap} was derived for the ground state of the gas, it provides a good description of the OBDM of 1D gases at the experimentally relevant temperatures and number of particles without the need of carrying out costly numerical calculations.

%%%%%%%%%%%%%%%%%%%%%%%%%%%%%%
%%%%%%%%%%%%%%%%%%%%%%%%%%%%

%%%%%%%%%%%%%%%%%%%%%%%%%%%%%%

\subsection{Soft-core bosons}\label{sec:LLvsQMC}

%@@@@@@@@@@@@@@@@@@@@@@@@@@@@@@@@@@@@@@
\begin{figure}[!t]
\includegraphics[width=\columnwidth]{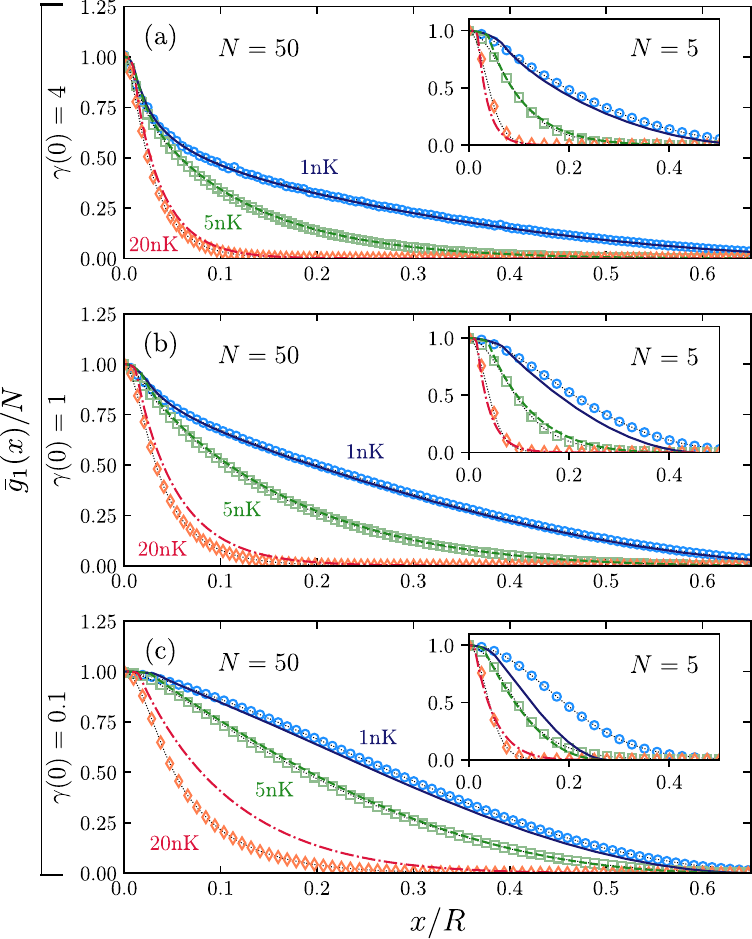}
\vspace{-0.5cm}
\caption{Average OBDM [see Eq.~\eqref{eq:avg-OBDM}] of a gas of $^{162}$Dy atoms for different values of the contact interaction confined in a harmonic trap with frequency $36.4$Hz. The reduced coupling at the center of the trap is: (a) $\gamma(0)=4$, (b) 1, and (c) 0.1. In the main panels we show results for $N=50$ and in the insets for $N=5$. Symbols joined by thin dotted lines are used for the numerical results obtained using quantum Monte Carlo simulations (referred to in the text as the ``{\it exact results}'') and thick continuous lines are used for the minimum between the result from the method in Sec.~\ref{sec:nummethod} (the long-distance asymptotics) and the result from Eq.~\eqref{eq:short-trap} (the short-distance expansion), which is referred to in the text as the results within ``{\it our approach}''. We report results for three temperatures: $T=1,\,5,$ and 20 nK.}\label{fig:scb-g1}
\end{figure}
%@@@@@@@@@@@@@@@@@@@@@@@@@@@@@@@@@@@@@@
%@@@@@@@@@@@@@@@@@@@@@@@@@@@@@@@@@@@@@@
\begin{figure}[!t]
\includegraphics[width=\columnwidth]{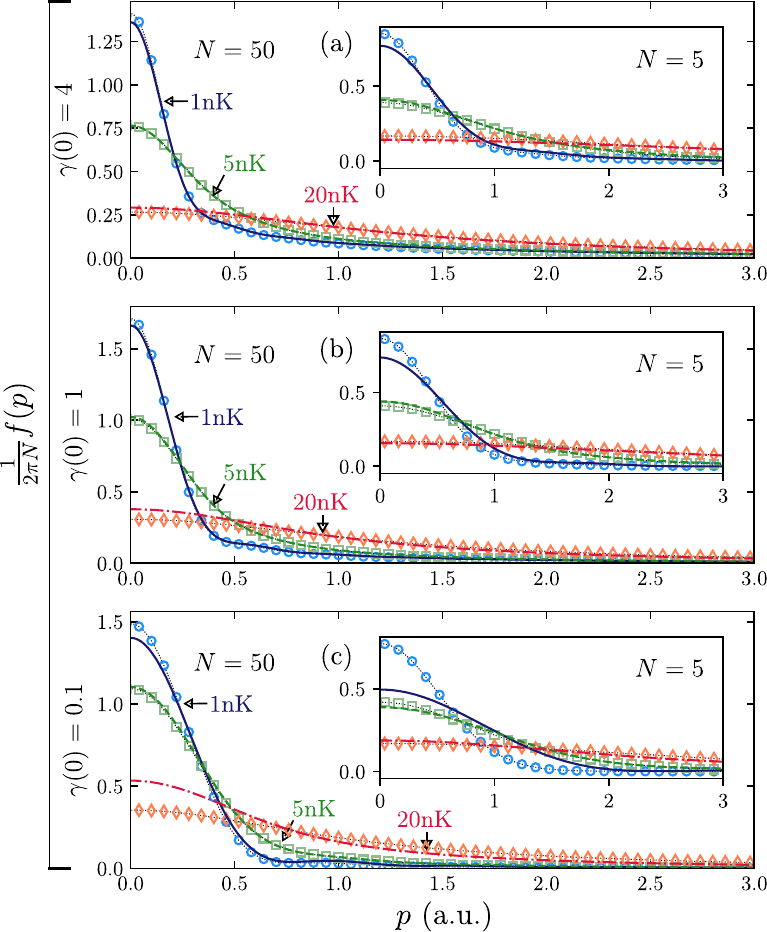}
\vspace{-0.5cm}
\caption{Same as Fig.~\ref{fig:scb-g1} but for the momentum distribution, which is obtained via a Fourier transform of the OBDM.}\label{fig:scb-md}
\end{figure}
%@@@@@@@@@@@@@@@@@@@@@@@@@@@@@@@@@@@@@@
In Figs.~\ref{fig:scb-g1} and~\ref{fig:scb-md}, we report the results for the average OBDM [see Eq.~\eqref{eq:avg-OBDM}] and the momentum distribution of soft-core bosons, respectively, with different strengths of the contact interactions. The main panels show results for $N=50$ and the insets show results for $N=5$. The behavior of differences between {\it our approach} and the {\it exact results} with increasing the temperature and/or decreasing the number of particles is qualitatively similar independently of the value of the contact interaction, which is reported in the figures in the form of the reduced coupling at the center of the trap, and is qualitatively similar to that discussed in detail in Sec.~\ref{sec:LLvsExactHCB} in the hard-core limit. Quantitatively, we find that the differences increase slowly as the strength of the contact interactions decreases, which is understandable as the effect of finite temperatures are enhanced as the interactions become weaker.

In Table~\ref{tab:soft}, we report the values of $\Delta$ [see Eq.~\eqref{eq:errorD}] obtained for different temperatures and strengths of the contact interactions. The differences are quantitatively similar to those obtained in the hard-core limit. Also, as expected from our previous discussions, $\Delta$ increases when the number of particles and the interaction coupling $\gamma(0)$ decrease. 

%@@@@@@@@@@@@@@@@@@@@@@@@@@@@@@@@@@@@@@
\begin{table}[!b]
\centering
\begin{tblr}{hlines,hline{1,Z} = 1pt,hline{2,3,4}=1pt,colspec={|r|cc |cc |cc|}}
   \SetCell[c=6]{c}          $\Delta$ for soft-core bosons  \\
   \hline
\SetCell[c=1]{c}$T$ (nK) & \SetCell[c=2]{c}   1 & & \SetCell[c=2]{c}  5 & & \SetCell[c=2]{c} 20 \\
$N$ & $50$ & $5$ & $50$ & $5$ & $50$ & $5$ \\
$\gamma(0)=4$  
& 0.032 & 0.137 & 0.021 &  0.084 & 0.05 & 0.219   \\
$\gamma(0)=1$
& 0.043 & 0.18 & 0.022 &  0.087 & 0.13 & 0.159   \\
$\gamma(0)=0.1$
& 0.087 & 0.443 & 0.029 &  0.079 & 0.287 & 0.138   \\
\end{tblr}
    \caption{Integrated difference $\Delta(p_{\max})$ in Eq.~\eqref{eq:errorD} for soft-core bosons for different temperatures/particle numbers (different columns/subcolumns), and different interaction couplings at the trap's center (different rows). We set $p_{\max}=8$. }
    \label{tab:soft}
\end{table}
%%%%%%%%%%%%%%%%%%%%%%%%%%%%%%%

\begin{figure}[!t]
    \centering
\includegraphics[width=.85\columnwidth]{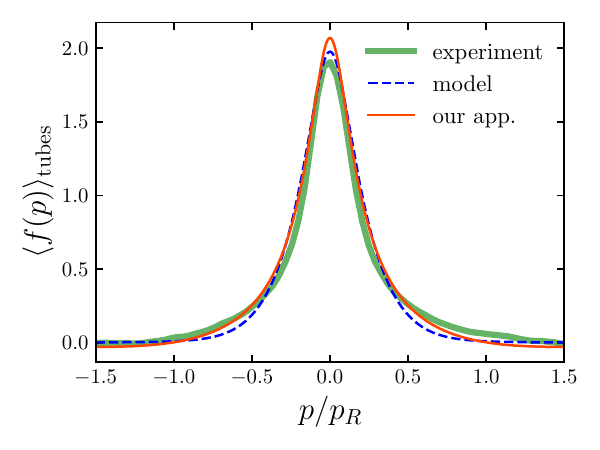}
\vspace{-0.4cm}
  \caption{Same as Fig.~\ref{fig:compare_exp_hard} but for 1D dysprosium gases in the soft-core regime with contact interaction strength $c = 8.5\,  \mu{\rm m}^{-1}$. The experimental data (thick solid line) and model results that use the same {\it exact approach} used here (dashed line), reported in Fig.~3(b) of Ref.~\cite{li2023dipolar}, are compared to our approach (thin solid line).}
    \label{fig:compare_exp_soft}
\end{figure}

In Fig.~\ref{fig:compare_exp_soft}, we compare experimental results for the momentum distribution for a contact interaction coupling $c=8.5\,\mu$m$^{-1}$ (corresponding to an estimated averaged value of $\gamma=6.7$~---~see Ref.~\cite{li2023dipolar} for details) and their modeling based on the {\it exact approach} used in this section, which were reported in Fig.~3(b) of Ref.~\cite{li2023dipolar}, to the result obtained using {\it our approach}. We find good agreement as in the hard-core limit, despite the fact that in the experiments $\approx$13\% of the particles are 1D gases with five or fewer particles, and that the average temperature of the gases is $\approx$17 nK.

%%%%%%%%%%%%%%%%%%%%%%%%%%%%%%
%%%%%%%%%%%%%%%%%%%%%%%%%%%%

\section{Conclusion}

\label{sec:conclusion}

We introduced a general approximate framework for estimating the one-body correlations and the momentum distributions of trapped 1D Bose gases at finite temperature and arbitrary interaction strengths. Our framework builds on results for the large-distance asymptotics of the OPDM derived using the inhomogeneous Luttinger liquid method~\cite{cazalilla2004bosonizing, dubail2017conformal, brun2017one, brun2018inhomogeneous, Bastianello_2020, scopa2020one}, combined with established short-distance expansions for the Bose gas~\cite{gangardt2004universal, olshanii2003short, Olshanii_2017}. Analytical results for the long-distance asymptotics were discussed in the asymptotic limits of weak repulsion in a harmonic trap (Eq.~\eqref{eq:OBDM-bogo-final}, derived previously in Refs.~\cite{Petrov2004, Citro2008}), and strong repulsion in an arbitrary trapping potential [Eq.~\eqref{eq:g1-tg-thermal}]. At finite interaction strength, the finite-temperature Green's functions of the inhomogeneous Luttinger liquid that enters the general formula for the OPDM [Eq.~\eqref{eq:main-OBDM-trap}] needs to be computed numerically. We explained in detail the numerical method for obtaining these Green’s function in Sec.~\ref{sec:nummethod}. The calculations are then carried out using Eq.~\eqref{eq:main-OBDM-trap} to evaluate the long-distance behavior of the OPDM and Eq.~\eqref{eq:short-trap} for the short-distance behavior. This allows to estimate the momentum distribution of the gas after carrying out a Fourier transform.

We benchmarked our framework against exact numerical calculations in the hard-core limit and quantum Monte Carlo simulations for finite repulsive interaction strengths. We also showed that our framework provides a good description of recent experimental results. The latter is remarkable because of the small numbers of particles in many of the 1D gases and the relatively high temperatures involved. This is promising as our approach is computationally inexpensive (our calculations can be carried out within a few minutes on a laptop) while the quantum Monte Carlo simulations require hundreds of CPU hours and need to be carried out in computing clusters. 

A crucial advantage of the approach we introduced here, which we plan to exploit next, is that unlike quantum Monte Carlo simulations it is not restricted to equilibrium situations. Our next goal is to determine how accurately it can describe the evolution of the momentum distribution following trap quenches as those studied in Refs.~\cite{malvania2021generalized, li2023dipolar}. Computing the dynamics of such momentum distributions is out of the reach of any existing theoretical method ---except in the asymptotic regime of hard-core bosons where exact results are available~\cite{Scopa_2023} and numerical calculations have already been compared to the experimental results~\cite{wilson2020observation, yang2023phantom}.\\

{\it Acknowledgements.}~---~We acknowledge P. Ruggiero and  A. Bastianello for discussions and joint work on closely related topics. We thank G. De Rosi for useful remarks on the paper. We acknowledge support from `` Lorraine Universit\'e d’Excellence" program (AT); ERC Consolidator Grant 771536 (NEMO) (SS and PC); ERC Starting Grant 101042293 (HEPIQ) (SS); MSCA Grant 101103348 (GENESYS) (SS); Agence Nationale de la Recherche through ANR-20-CE30-0017-02 project ‘QUADY’ (JD) and ANR-22-CE30-0004-01 project ‘UNIOPEN’ (JD); Dodge Family Postdoc Fellowship at the University of Oklahoma (YZ); and the National Science Foundation under Grant No.~PHY-2309146 (MR). \emph{This work has been partially funded by the European Union. Views and opinions expressed are however those of the author(s) only and do not necessarily reflect those of the European Union or the European Commission. Neither the European Union nor the European Commission can
be held responsible for them.}
\appendix

\section{Calculation of the nonuniversal amplitudes}\label{app:nonuniv-coef}

Here we summarize the procedure used to calculate the nonuniversal amplitudes $B_m$ entering in the long-distance asymptotics of the OBDM in Eq.~\eqref{eq:LL_exp}. We refer to, e.g., Refs.~\cite{shashi2011nonuniversal, shashi2012exact, brun2018inhomogeneous, scopa2020one} for detailed studies of these amplitudes.

\begin{figure} [htb]
\includegraphics[width=\columnwidth]{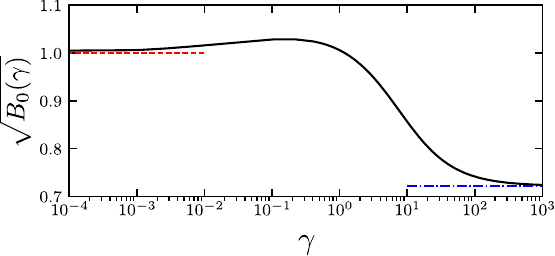}
\vspace{-0.4cm}
\caption{Nonuniversal field amplitude $\sqrt{B_0(\gamma)}$ as a function of $\gamma$. The
horizontal lines depict the asymptotic results $B_0(\gamma\to 1)\to 1$~\cite{brun2018inhomogeneous} (dashed line) and $B_0(\infty)\simeq 0.5213$~\cite{vaidya1979one} (dotted-dashed line).}
    \label{fig:nonuniv-ampl}
\end{figure}

By exploiting the operator-state correspondence between vertex operators in the underlying effective field theory and the low-energy excited states of the microscopic model, we write the field amplitude for a homogeneous system of $N$ bosons on a ring of size $L$ as~\cite{brun2018inhomogeneous, scopa2020one}
\begin{align}
    \label{eq:formfact}
\sqrt{B_{m}} &=  \lim_{N,L \rightarrow \infty} \left(\frac{L}{2 \pi}\right)^{\Delta_m} \times \nonumber \\
&    \frac{\left| \bra{\{J_j^{(m)}\}^{N-1}_{j = 1}} \hat{\Psi}(0) \ket{\{ I_i\}^{N}_{i = 1}} \right|}{ \sqrt{\braket{\{J_j^{(m)}\}^{N-1}_{j = 1}}} \sqrt{\braket{\{I_i\}^{N}_{i = 1}}}},
\end{align}
where the matrix elements can be efficiently computed using the determinant formula of Ref.~\cite{slavnov1989}. The state $\ket{\{I_i\}^{N}_{i = 1}}$ is the ground state of the gas, specified by the set of Bethe integers 
\be\label{eq:bethe-int-gs}
I_i =   -\frac{N+1}{2} + i,   \qquad    i = 1,\dots,N, 
\ee
while $\ket{\{J_j^{(m)}\}^{N-1}_{j = 1}}$ is a low-energy excited state corresponding to the Bethe integers
\be\label{eq:bethe-int-umk}
J_j^{(m)} = -\frac{N}{2} + j+  m,  \qquad   j = 1,\dots,N-1.
\ee
We recall that the set of rapidities specifying the Bethe states above is obtained from Eqs.~\eqref{eq:bethe-int-gs} and~\eqref{eq:bethe-int-umk} by solving the Bethe equations~\eqref{eq:Betheeq}. Finally, the nonuniversal field amplitude \eqref{eq:formfact}  requires taking the thermodynamic limit $N, L\to\infty$ at fixed density $n=N/L$ and interaction $c$, i.e., at fixed reduced coupling $\gamma$. In practice, one can compute the amplitude for fixed values of $N$ and $L$ and extrapolate the thermodynamic limit with a polynomial fit in $1/N$. The result of this procedure is shown in Fig.~\ref{fig:nonuniv-ampl} for the leading term $m=0$.

%@@@@@@@@@@@@@@@@@@@
%@@@@@@@@@@@@@@@@@@@

\section{Analytical expression for the Green's function of trapped hard-core bosons at finite temperature}\label{app:cft-TG-details}

In this Appendix we derive Eq.~(\ref{eq:theta_corr_annulus}). We start from the mode decomposition of the Luttinger liquid Hamiltonian at $K=1$ with Dirichlet boundary conditions on both sides [i.e. $\hat{\phi}(s=0) = \hat{\phi}(s=1) = 0$],
\begin{eqnarray}
\nonumber \hat{\mathscr{H}}_\text{inh}  &=&  \int_0^1 \frac{ds}{2\pi{\cal T}}  \left[ (\partial_s \hat{\theta} )^2 +  (\partial_s \hat{\phi} )^2  \right]  = \frac{\pi}{\mathcal{T}}  \sum_{n \geq 1}  [ \hat{a}^\dagger_n \hat{a}_n  + \frac{1}{2} ] ,
\end{eqnarray}
where $\mathcal{T}$ is the time needed for a gapless excitation to travel from the left to the right boundary, and
\begin{eqnarray}
\hat{\phi}(s) &=& \sum_{n \geq 1 } \frac{1}{n} \sin ( \pi n s ) (\hat{a}^\dagger_n + \hat{a}_n)\, , \\
 \hat{\theta}(s) &=& \sum_{n \geq 1 } \frac{1}{i \, n}  \cos ( \pi n s) (\hat{a}^\dagger_n - \hat{a}_n)\, ,
\end{eqnarray}
with the canonical commutation relation 
\begin{equation}
[\hat{a}_n, \hat{a}_m^\dagger] \, = \, n \, \delta_{n,m}\, ,
\end{equation}
such that $[\hat{\phi}(s) , \frac{1}{\pi} \partial \hat{\theta}(s') ] =  [\hat{\phi}(s) ,\hat{\Pi}(s')  ] = i \delta(s-s')$ for $s,s'\in(0,1)$. The finite-temperature Green's function is
\begin{eqnarray*}
	\left< \hat{\theta} (s) \hat{\theta}(s') \right>_\beta  
	&= &  -\sum_{n} \frac{ \cos ( \pi n s )  \cos ( \pi n s' )}{n^2}  \left< (\hat{a}^\dagger_n - \hat{a}_n)^2  \right>_\beta  \\
	&= & 2 \sum_{n} \frac{  \cos ( \pi n s ) \cos ( \pi n  s' )}{n^2} \left< \hat{a}^\dagger_n  \hat{a}_n + \frac{n}{2} \right>_\beta  .
\end{eqnarray*}
We then use
\begin{equation}
    \left< \hat{a}^\dagger_n \hat{a}_n \right>_\beta \, = \, \frac{{\rm tr} \left[ e^{- \beta H} \hat{a}^\dagger_n \hat{a}_n \right] }{ {\rm tr} \left[ e^{- \beta H} \right] } \, = \, n \, \frac{ \sum_{p \geq 0} p q^{n p} }{ \sum_{p \geq 0} q^{n p} }\, ,
\end{equation}
with
\begin{equation}
    \label{eq:q_app}
    q =  \exp\left(- \pi \beta/\mathcal{T}\right)  ,
\end{equation}
and the following identity
\begin{eqnarray*}
	&& \sum_{n \geq 1} \frac{1}{n} u^n 	\frac{ \sum_{p \geq 0} (p + \frac{1}{2} ) q^{n p} }{ \sum_{p \geq 0} q^{n p} }  = \frac{1}{2} \sum_{n \geq 1} \frac{1}{n} u^n    \frac{1+q^n}{1-q^n}  \\
	&& \qquad = \frac{1}{2} \sum_{p \geq 0} \sum_{n \geq 1} \frac{1}{n} u^n  (1+q^n) q^{p n} \\
	&& \qquad = \frac{1}{2} \sum_{p \geq 0} - \log \left( 1 - u q^p \right)  - \log \left( 1 - u q^{p+1} \right)   \\	
	% && =   \frac{1}{2} \log \left( 1 - u \right)  - \log \left( \prod_{p \geq 0} (1 - u q^p) \right) \\	
	&& \qquad =  - \log \left[ (u; q)_\infty \right]\, + \,  \frac{1}{2} \log \left( 1 - u \right) ,
\end{eqnarray*}
where $
(u; q)_\infty  \, = \, \prod_{p \geq 0} (1 - u q^p) $ is known as the `$q$-Pochhammer symbol'. This leads to 
\begin{eqnarray*}
\langle \hat{\theta}(s) \hat{\theta}(s') \rangle_{\beta} &=&  - \frac{1}{2} \log \left[ \left( e^{i \pi (s - s')}, q \right)_\infty \left( e^{-i \pi (s - s')}, q \right)_\infty \right]  \nonumber \\&& - \frac{1}{2} \log \left[ \left( e^{i \pi (s + s')}, q \right)_\infty \left( e^{-i \pi (s + s')}, q \right)_\infty \right]  \nonumber \\
&&+  \frac{1}{4} \log \left[ \left( 1- e^{i \pi (s - s')}\right) \left( 1- e^{-i \pi (s - s')}\right) \right]  \nonumber \\ &&+ \frac{1}{4} \log \left[ \left( 1- e^{i \pi (s + s')}\right) \left( 1- e^{-i \pi (s + s')}\right)   \right] .
\end{eqnarray*}
Reorganizing the terms in the infinite products, one arrives at
\begin{align}
    \label{eq:phiphi_product_app}
 &\langle\hat{\theta}(s) \hat{\theta}(s') \rangle_\beta= \nonumber\\[4pt]
 &\quad  - \frac{1}{4} \log \left[ \prod_{n \in \mathbb{Z}} \left| \frac{ \sin \frac{\pi (s-s' - i n \frac{\beta}{\mathcal{T}} )}{2}  \sin \frac{\pi (s+s' - i n \frac{\beta}{\mathcal{T}} )}{2} }{ \left( \sin \frac{\pi (s- i n \frac{\beta}{\mathcal{T}} )}{2}  \sin \frac{\pi (s'- i n \frac{\beta}{\mathcal{T}} )}{2} \right)^2 } \right|^2 \right] ,
\end{align}
where one can see the periodicity of the result under $s \rightarrow s + i \frac{\beta}{\mathcal{T}} $. This result can be rewritten in terms of the Jacobi's $\theta$ function $\vartheta_1(z|\tau)$ [defined in Eq.~\eqref{eq:jacobi-theta} of the main text],
\begin{equation}
  \left< \hat{\theta}(s) \hat{\theta}(s') \right>_\beta  = - \frac{1}{2}\log{ \left| \frac{\vartheta_1 \!\left( \frac{s - s'}{2} \left|\frac{i\beta}{2 \Ltilde } \!\right. \right) \vartheta_1 \!\left( \frac{s + s'}{2} \left| \frac{i\beta}{2 \Ltilde } \!\right.\right)}{\left[\partial_s \vartheta_1 \!\left( 0 \left|\frac{i\beta}{2 \Ltilde } \!\right. \right) \right]^2 } \right| } ,
\label{eq:theta_corr_annulus_app}
\end{equation}
which is Eq.~(\ref{eq:theta_corr_annulus}) of the main text. A simple way to check that Eqs.~(\ref{eq:phiphi_product_app}) and (\ref{eq:theta_corr_annulus_app}) are equivalent is to do a series expansion in the parameter $q$ [defined in Eq.~(\ref{eq:q_app})] for both expressions and check that the coefficients of the expansions match at all orders.

We note that a similar calculation leads to an analogous result for the $\phi$-$\phi$ correlation function,
\begin{eqnarray}
    \label{eq:phi_corr_annulus_app}
\nonumber	\left< \hat{\phi}(s) \hat{\phi}(s') \right>_\beta  
	&=&  - \frac{1}{4} \log \left( \prod_{n \in \mathbb{Z}} \left| \frac{ \sin \frac{\pi(s-s' - i n \frac{\beta}{\Ltilde} )}{2}  }{ \sin \frac{\pi(s+s' - i n \frac{\beta}{\Ltilde})}{2}   } \right|^2 \right)  \\
 &=&  - \frac{1}{2} \ln{ \left| \frac{\vartheta_1(\frac{{ x - x'}}{2}|i\beta/2)}{\vartheta_1(\frac{{ x + x'}}{2}|i\beta/2)}  \right| }.
\end{eqnarray}
Alternatively, Eqs.~(\ref{eq:theta_corr_annulus_app}) and~(\ref{eq:phi_corr_annulus_app}) can be obtained by using the correlation function of the 2D massless free boson theory on a torus, which reads (see, e.g., Eq.~(12.142) in the textbook~\cite{DiFrancesco:1997nk})
\begin{equation}
    \langle \phi (z,\bar{z}) \phi(0,0) \rangle_{\rm torus} = - \frac{1}{2} \log{ \left| \frac{\vartheta_1(z|i\tau )}{\partial_{z} \vartheta_1( 0 | i \tau )} e^{- \pi \frac{ ({\rm Im} \, z)^2}{\tau}} \right| }.
    \label{eq:phi_corr_torus}
\end{equation}
Then, applying the method of images (see, e.g., Chapter 9 in Ref.~\cite{DiFrancesco:1997nk} or Refs.~\cite{cazalilla2004bosonizing, brun2017one}) to construct the two-point function on an annulus with Dirichlet boundary conditions on both sides, one arrives at Eq.~(\ref{eq:phi_corr_annulus_app}). The same exercise with Neumann boundary conditions leads to Eq.~(\ref{eq:theta_corr_annulus_app}).

\newpage
 
\bibliography{finitetemperature}
\end{document}